\documentclass[twocolumn,preprintnumbers,superscriptaddress,amsmath,amssymb,10pt,a4paper]{revtex4-1}

\usepackage{geometry}
\usepackage{mathtools}
\usepackage{graphicx}
\usepackage{dcolumn}
\usepackage{bm}
\usepackage{makecell}
\usepackage{xcolor}
\usepackage{enumerate}
\usepackage[caption=false]{subfig}
\usepackage{float}
\usepackage[unicode=true,pdfusetitle,
 bookmarks=true,bookmarksnumbered=false,bookmarksopen=false,
 breaklinks=false,pdfborder={0 0 1},backref=false,colorlinks=true]
 {hyperref}
\hypersetup{linkcolor=blue,urlcolor=blue,citecolor=blue}

\newcommand\numberthis{\addtocounter{equation}{1}\tag{\theequation}}

\newcommand{\be}{\begin{equation}}
\newcommand{\ee}{\end{equation}}
\newcommand{\bd}{\begin{displaymath}}
\newcommand{\ed}{\end{displaymath}}
\newcommand{\BE}{\begin{eqnarray}}
\newcommand{\EE}{\end{eqnarray}}

\newcommand{\bv}{\ensuremath{\mathbf{v}}}

\newcommand{\bn}{\ensuremath{\mathbf{n}}}

\newcommand{\bA}{\ensuremath{\mathbf{A}}}

\newcommand{\brho}{{\mbox{\boldmath $\rho$}}}
\newcommand{\bnu}{{\mbox{\boldmath $\nu$}}}

\newcommand{\avg}[1]{\left\langle{#1}\right\rangle}

\DeclarePairedDelimiter\abs{\lvert}{\rvert}%

\begin{document}

\title{Beyond the adiabatic limit in systems with fast environments:\\ a $\tau$-leaping algorithm}

 \author{Ernesto Berr\'ios-Caro}
\email{ernesto.berrios@postgrad.manchester.ac.uk}
\affiliation{Theoretical Physics, Department of Physics and Astronomy, School of Natural Sciences, Faculty of Science and Engineering, The University of Manchester, Manchester M13 9PL, United Kingdom}

\author{Tobias Galla}
\email{tobias.galla@ifisc.uib-csic.es}
\affiliation{Theoretical Physics, Department of Physics and Astronomy, School of Natural Sciences, Faculty of Science and Engineering, The University of Manchester, Manchester M13 9PL, United Kingdom}
\affiliation{Instituto de F\'isica Interdisciplinar y Sistemas Complejos, IFISC (CSIC-UIB), Campus Universitat Illes Balears, E-07122 Palma de Mallorca, Spain}

\date{\today}

\begin{abstract}
We propose a $\tau$-leaping simulation algorithm for stochastic systems subject to fast environmental changes. Similar to conventional $\tau$-leaping the algorithm proceeds in discrete time steps, but as a principal addition it captures environmental noise beyond the adiabatic limit. The key idea is to treat the input rates for the $\tau$-leaping as (clipped) Gaussian random variables with first and second moments constructed from the environmental process. In this way, each step of the algorithm retains environmental stochasticity to sub-leading order in the time scale separation between system and environment.  We test the algorithm on several toy examples with discrete and continuous environmental states, and find good performance in the regime of fast environmental dynamics. At the same time, the algorithm requires significantly less computing time than full simulations of the combined system and environment. In this context we also discuss several methods for the simulation of stochastic population dynamics in time-varying environments with continuous states.
\end{abstract}
\maketitle

\section{Introduction}

The modelling of dynamical systems in biology and other disciplines necessarily requires simplifying assumptions and a level of coarse graining. If all processes we know about are included, then the model becomes so complicated that it cannot be simulated or analysed. Even if simulation or analysis is possible further study of such a model will rarely be enlightening. Excessive detail makes hard to identify the key mechanisms at work and to understand what model components are responsible for these mechanisms. At the same time, some element of realism must be maintained. The model must not be so stylised to miss the key ingredients and behaviour it is meant to capture. The principal challenge, therefore, is to find the right level of detail, given the intended purpose.

The choice between stochastic and deterministic modelling approaches is one aspect of this discussion. If more detailed stochastic models mark one end of the spectrum, then many traditional models in mathematical biology or chemistry sit at the opposite end. These models are often built on a small number of ordinary or partial differential equations (e.g. \cite{murray1,murray2}). This deterministic approach is valid if one can assume that the same initial conditions will always lead to the same outcome. For many applications involving very large systems this is a perfectly sensible approach. 

However, it is now also universally recognised that stochasticity in the time-evolution of many systems is key in shaping the outcome, see e.g. \cite{Goel2004,ewens,TraulsenHauert}. Consequently a number of analytical and computational methods has been developed for the study of stochastic systems. One focus is on systems with discrete interacting individuals. What these individuals represent depends on the context, they could be members of different species in population dynamics, individual animals or humans in models of an epidemic, or molecules in chemical reaction systems \cite{Goel2004,castellano,keeling,vankampen}.

One particular point of interest within this class of individual-based systems are models operating in a time-dependent environment. This environment is not part of the system proper, but its state has an effect on what happens in the system. In a model of a population of bacteria for example, the reproduction or death rates could depend on external conditions  such as the availability of nutrients or the presence of toxins \cite{acar2008stochastic,patra2015emergence}. In population dynamics, the carrying capacity could vary in time \cite{wienand,wienand2,taitelbaum2020}, and in epidemics the infection rate is subject to seasonal changes \cite{black2010stochastic}. The focus of our paper is on such individual-based models in time-varying external environments. 

Analytical approaches to stochastic systems with discrete individuals usually start from the chemical master equation. In limited cases direct solution is possible, for example using generating functions. However, this is the exception, and a number of approximation schemes have consequently been developed. These include Kramers--Moyal and system-size expansions, leading to Fokker--Planck equations and descriptions in terms of stochastic differential equations \cite{vankampen,gardiner}. These schemes sacrifice the granularity of a discrete-agent system, and instead describe the dynamics in terms of continuous densities. This approach can be successful in particular for large populations. Any particular event then only results in a small change in the composition of the population relative to its size. Individual-based approaches and descriptions based on deterministic differential equations been extended to models of population dynamics in switching environments, for a selection of work see \cite{kussell2005phenotypic,kepler2001stochasticity,thattai2004stochastic,swain2002intrinsic,assaf2013extrinsic,duncan2015noise,assaf2,ashcroft2014fixation,wienand,west2018,assaf2008population,hufton2019model}.  

There are however situations in which one would rather avoid giving up the discrete nature of the population. For example, granularity is crucial for extinction processes (the number of individuals of the species about to go extinct is small by definition). In other situations the population may not be large enough to warrant a description in terms of continuous densities. For example, copy numbers in genetic circuits can be of the order of tens to hundreds (see e.g. \cite{eldar2010}), and it is difficult to justify a continuum limit. It then becomes necessary to carry out numerical simulations of the discrete individual-based process. The method of choice is the Gillespie algorithm \citep{gillespie1976general, gillespie1977exact}, generating a statistically accurate ensemble of sample paths of the continuous-time dynamics. 

In most applications the rate of events scales with the size of the population so that each individual experiences an ${\cal O}(1)$ number of reactions per unit time. The Gillespie method then runs into difficulties when the population is large, and with it the number of reactions per unit time. The computational cost of generating sample paths to up the desirable end point can then become very high. Similarly, a time scale separation between the dynamics in the population and the environment may make simulations challenging. If the environment is very fast compared to the population, a significant number of environmental events needs to be executed between events in the population. This aggravates the above limitations for large populations, and simulations can become problematic even for intermediate population sizes.  One possible approach to this consists of assuming that the environment is `infinitely' fast compared to the population. This is known as quasi steady state approximation \cite{QSSA1,QSSA2},  or the `adiabatic limit' \cite{lin2018efficient,hufton2019model}. For related work see also \cite{Newby2010,ashcroft2014fixation, Bressloff2016,Bressloff2017,Bressloff2017a}. If this limit is taken then the environmental dynamics can be `averaged out', and effective reaction rates can be used for the population. While computationally convenient, this approach discards any stochasticity from the environmental process. This sets another limitation, in particular when it is not valid to assume that the environment is infinitely fast compared to the population.  

The objective of this work is therefore to design and test an algorithm for systems with fast environmental dynamics, but which also captures some elements of the environmental noise. We call this discrete-time algorithm $\tau$FE -- $\tau$-leaping for fast environments. It is built on the ideas of the conventional $\tau$-leaping algorithm \citep{Gillespie_2001}, but with modifications such as to preserve elements of the stochasticity of the environmental process. To do this, we assume that the environment is fast compared to the population, but not infinitely fast. More precisely, in each step of the algorithm we take into account sub-leading contributions in the time-scale separation.  

The key new element of our algorithm is how we deal with the environment. We do not take the adiabatic limit, instead we treat the reaction rates in the population as random variables during each step. The rates are drawn from a distribution at the beginning of each step, and then remain fixed during the time step. The distribution of rates can change from one step to the next, and is constructed to reflect statistical features of the original environmental dynamics. 

Each step of the $\tau$FE algorithm consists of two parts: First a realisation of reaction rates is drawn from the appropriate distribution. Then a conventional $\tau$-leaping step is carried out with these rates. The core of our paper consists of the construction of the `appropriate distribution' for the reaction rates. These ideas were introduced in a previous work \cite{hufton2019classical} for a simple case of a two-species birth-death process in an environment which can take two discrete states. In the present paper we develop this further. We develop and test a more general algorithm for environments with more than two discrete states. As we will show, the algorithm can also be extended to continuous environmental dynamics. 

The remainder of the paper is organised as follows. In Sec.~\ref{sec:model} we describe the general setup of the type of system we simulate. We also outline the general principles of the $\tau$FE algorithm. In Sec.~\ref{sec:algo} we then make the necessary preparations for the introduction of the algorithm. In particular, we compute the statistics of reaction rates which are fed into the conventional $\tau$-leaping step. We then describe the algorithm in detail. In Sec.~\ref{sec:discrete}, we test the $\tau$FE algorithm in different models with discrete environmental states. In Sec.~\ref{sec:continuous}, we  then describe how the $\tau$FE algorithm can be used when the environment takes continuous states. Specifically, we consider an Ornstein-Uhlenbeck process. In this context we also describe how known algorithms can be adapted to simulate continuous environments. Finally, we provide a discussion of our results and overall conclusions in Sec.~\ref{sec:conclusions}. 

\section{Model setup and general principles of the algorithm}\label{sec:model}
\subsection{Model definitions and notation}\label{sec:def}
We look at systems composed of discrete individuals. We will refer to this synonymously as the `system proper', or `the population'. Each of the individuals is of one of $S$ species (or types), labelled $i=1,\dots,S$. We write $n_i$ for the number of individuals of species $i$ in the population, and $\bn=(n_1,\dots,n_S)$. The system evolves in an external environment, whose state we write as $\sigma$. These states are time dependent, and can either take discrete values or be continuous.

The dynamics in the population proceeds through reactions $r=1,\dots,R$. Each of these reactions converts a number of individuals from one type into another. Time in the model is continuous, and we assume that the dynamics is Markovian. We then write $R_{r,\sigma}(\bn)$ for the rate of reaction $r$ if the environment is in state $\sigma$ and the population in state $\bn$. The stoichiometric coefficient $\nu_{r,i}$ indicates how the number of individuals of type $i$ changes when a reaction of type $r$ occurs. Each $\nu_{r,i}$ is an integer, which can be negative, zero, or positive. We write $\bnu_r=(\nu_{r,1},\dots,\nu_{r,S})$. The rates $R_{r,\sigma}(\bn)$ and the stoichiometric coefficients fully specify the dynamics of the population when the environment is in state $\sigma$.

The state of the environment undergoes a Markovian stochastic process, governed by a master equation if states are discrete or by a stochastic differential equation in the case of continuous environmental states. These dynamics can depend on the state of the population $\bn$. If the environmental states are discrete we write $q_{\sigma\to\sigma'}(\tau)$ for the probability of finding the environment in state $\sigma'$ at a particular point in time, given that $\tau$ units of time earlier it was in state $\sigma$. If the environment is continuos then $q_{\sigma\to\sigma'}(\tau)$ is a probability density for $\sigma'$ (at given $\sigma$). We call  $q_{\sigma\to\sigma'}(\tau)$ the transition kernel of the environmental process. We write $\brho^*$ for the stationary distribution of the environmental dynamics. For discrete environmental states the entries $\rho^*_\sigma$ denote the probability of finding the system in state $\sigma$ in the stationary state. For continuous environments $\rho^*_\sigma$ is the stationary probability density for $\sigma$.

\subsection{General principles of the $\tau$-leaping algorithm for systems in fast environments}
A conventional reaction system (without external environment) is governed by a chemical master equation of the form
\begin{align*}
& \frac{\mathrm{d}}{\mathrm{d}t} P(\bn,t)=	\\
& \sum_r \big[R_r(\bn-\bnu_r)P(\bn-\bnu_r,t) -R_r(\bn)P(\bn,t\big)]. \numberthis
\end{align*}
The notation is as in Sec.~\ref{sec:def}, the only difference is that there is no subscript $\sigma$, as there is no environment. Sample paths entail events (reactions) which can occur at any point in continuous time, separated by exponentially distributed random waiting times. In each such event the state of the system $\bn$ changes, and accordingly the reaction rates $R_r(\bn)$ can also change. Sample paths can be generated for example using the celebrated Gillespie algorithm \citep{gillespie1976general, gillespie1977exact}.

The $\tau$-leaping algorithm for such conventional reaction systems is built around the idea of keeping reaction rates constant over finite time steps of length $\tau$ \cite{Gillespie_2001}. That is to say, if the state of the population is $\bn$ at time $t$, then the assumption is made that this state $\bn$ and the rates $R_r(\bn)$ do not change until the end of the time step. The algorithm does not account for potential changes of the rates as individual reactions occur, and instead directly `leaps' to time $t+\tau$. This is justified provided the so-called `leap condition' is fulfilled \cite{Gillespie_2001}: broadly speaking the time step $\tau$ must be sufficiently small so that the state $\bn$ in the continuous-time system does not change significantly in a time interval of length $\tau$.

Making the approximation of constant $\bn$ in the time interval from $t$ to $t+\tau$, the number of reactions of type $r$ that fire in this interval follows a Poissonian distribution with parameter $\tau R_r(\bn)$. Accordingly, realisations of Poissonian random variables $m_1,\dots,m_R$ are drawn, and the corresponding numbers of each reactions are executed simultaneously. This generates a new state $\bn'$ at time $t+\tau$, with entries $n_i'=n_i+\sum_{r=1}^R m_r \nu_{i,r}$. The process then repeats with updated rates $R_r(\bn')$.

The idea of the $\tau$-leaping algorithm we introduce for systems in external environments is similar. As in the conventional algorithm we discretise time, and keep the composition of the population $\bn$ fixed during each iteration. It is only updated at the end of each step. From now on we use $\Delta t$ for the duration of a step instead of $\tau$. 

The difference to the conventional case is the external environment. If the environmental state space is discrete then switches of the environment can in principle be simulated in continuous time along with the other reactions (using Gillespie algorithm \citep{gillespie1976general, gillespie1977exact}). They can also be dealt with by means of the conventional $\tau$-leaping algorithm, again along with the other reactions. These are natural simulation approaches when the environment operates on a similar time scale as the reactions in the population. Not much can then be gained by distinguishing between environmental processes and the dynamics in the system proper.

If the environment is infinitely fast compared to the reactions in the population, then the environment reaches stationarity on very short time scales. One can average over environmental states, see for example  \cite{QSSA1,QSSA2,Newby2010,Bressloff2014,hufton2019classical,hufton2019model}. If the environment is discrete, for example, we can use average rates
\be\label{eq:rstar}
R_{r}^*(\bn)\equiv\sum_\sigma \rho^*_\sigma R_{r,\sigma}(\bn).
\ee
In the case of continuous environments the sum is to be replaced with an integral. These rates are functions of $\bn$ only, the environmental process has been averaged out. Noise from the environmental process plays no role in the dynamics if these average rates are used. This corresponds to making a quasi-stationary state approximation for the fast-moving environment \cite{QSSA1,QSSA2}.

The aim of this paper is to go beyond this adiabatic limit, and to construct a $\tau$-leaping algorithm which captures some elements of extrinsic noise. We focus on the limit of a fast, but not infinitely fast environmental dynamics. 

Broadly speaking the $\tau$FE algorithm is constructed around the idea of treating the reaction rates $R_r(\bn)$ as stochastic variables in each discrete time step. These random variables represent the rates one obtains when averaging the environmental process over the time step $\Delta t$. Assuming that the rate of change of the environment is finite these average rates will remain stochastic. In the limit of infinitely fast environments the deterministic limit in Eq.~(\ref{eq:rstar}) is recovered, and there is no stochasticity from the environment. 

To construct the random reaction rates for each step, we make an approximation: we use a Gaussian distribution for the rates, with means as in Eq.~(\ref{eq:rstar}) and with variances and correlations derived from the original combined process of the population and environment. We describe this in detail in the next section.

\section{Construction of the $\tau$FE algorithm for systems with discrete environments}
\label{sec:algo}

\subsection{Preliminary analysis of the environmental process}\label{sec:prelim}

Here we assume the environmental states are discrete, $\sigma\in\{1,\dots, M\}$. The dynamics of the environment is governed by the rates $\lambda A_{\sigma\to\sigma'}(\bn)$ for transitions from $\sigma$ to $\sigma'$. The factor $\lambda$ is introduced to control the time-scale separation between reactions in the population and the switching of the environment. We use the notation $\bA(\bn)$ for the $M\times M$ matrix with elements $A_{\sigma\to\sigma'}(\bn)$. We also set $A_{\sigma\to\sigma}(\bn)=-\sum_{\sigma'\neq \sigma} A_{\sigma\to\sigma'}(\bn)$. The combined dynamics of population and environment are then described by the master equation
\begin{align*}
&\frac{\mathrm{d}}{\mathrm{d}t}P(\bn,\sigma,t)=  \\
&\sum_r \big[R_{r,\sigma}(\bn-\bnu_r)P(\bn-\bnu_r,\sigma,t) -R_{r,\sigma}(\bn)P(\bn,\sigma,t)\big] \\
& + \lambda\sum_{\sigma'} \big[A_{\sigma'\to\sigma}(\bn)P(\bn,\sigma',t) -A_{\sigma\to\sigma'}(\bn)P(\bn,\sigma,t)\big]. \numberthis
\end{align*}

The rates $\lambda A_{\sigma\to\sigma'}(\bn)$ can depend on the state of the population, $\bn$. This means that $\bn$ and $\sigma$ do not necessarily evolve in time independently. However, as mentioned above the state $\bn$ of the system is kept constant during each $\tau$-leaping step. This in turn means that the transition rates for the environment also remain constant during each step. 

We now focus on one such time step, starting at time $t$ and ending at $t+\Delta t$. We assume that $\bn$ remains constant during this time interval. For the remainder of Sec.~\ref{sec:prelim} we suppress the potential dependence of $\bA$ on $\bn$, although it is always implied. We write $\rho_\sigma(t')$ for the probability that the environment is in state $\sigma$ at time $t\in [t,t+\Delta t]$. We then have the master equation
\be\label{eq:masterenv}
\frac{\mathrm{d} \brho }{\mathrm{d}t'}= \lambda\bA \brho
\ee
for the environmental dynamics. The stationary distribution $\brho^*$ for the environment is the solution of $\bA\brho^*=0$. If $\bA$ depends on $\bn$, then $\rho^*$ will also be a function of $\bn$. Assuming that the environmental process is irreducible this stationary distribution is unique for any one $\bn$.

The stochastic matrix $\bA$ has one zero eigenvalue, which we write as $\mu_1=0$. The remaining eigenvalues are denoted by $\mu_2,\dots,\mu_M$. We then have $\mu_2,\dots,\mu_M<0$. The corresponding (right) eigenvalues are written as $\bv_1=\brho^*$ (the eigenvector corresponding to eigenvalue $0$), and $\bv_2,\dots,\bv_M$ respectively for the remaining eigenvectors. These are all understood to be column vectors of length $M$.

We note that the general solution of Eq.~(\ref{eq:masterenv}) can be written in the form
\be
\brho(t')=\brho^*+\sum_{\ell=2}^M c_\ell e^{\lambda\mu_\ell (t'-t)} \bv_\ell,
\label{eq:solution-rho}
\ee
with coefficients $c_\ell$ determined by the initial condition at the beginning of the time step $t'=t$. More precisely these coefficients can be obtained from the linear system
\be
\sum_{\ell=2}^M c_\ell \bv_\ell=\brho(t)-\brho^*.
\label{eq:coeff}
\ee
We remark that there are $M-1$ coefficients, $c_\ell$ ($\ell=2,\dots,M$). The system in Eq.~(\ref{eq:coeff}) technically consists of $M$ equations, but these are not independent due to normalisation of the probabilities on the right-hand side.

Calculating the probability $q_{\sigma\to\sigma'}(\Delta t)$ to find the environment in state $\sigma'$ at the end of the time step if it was in $\sigma$ at the beginning of the step is now mainly a matter of computing the coefficients $c_\ell$. We write $c_{\ell,\sigma}$ for the value the coefficient $c_\ell$ takes when $\rho_{\sigma'}(t)=\delta_{\sigma',\sigma}$ (i.e., when the system starts in state $\sigma$ at the beginning of the step).

We then have
\be\label{eq:q}
q_{\sigma\to\sigma'}(\Delta t)=\rho^*_{\sigma'}+\sum_{\ell=2}^M c_{\ell,\sigma} e^{\lambda\mu_\ell \Delta t} v_{\ell,\sigma'},
\ee
where $v_{\ell,\sigma'}$ is the $\sigma'$-entry of the eigenvector $\bv_\ell$ of $\bA$.

If the matrix $\bA$ depends on the population state $\bn$, the parameters $\mu_\ell, v_{\ell,\sigma'},$ and $c_{\ell,\sigma}$ can also depend on $\bn$. For simplicity of notation we have not included this potential dependence in the above equations.

\subsection{Time-averaged reaction rates as random variables}

The $\tau$-leaping algorithm proceeds in discrete time intervals of length $\Delta t$. We continue to focus on one such interval $[t,t+\Delta t]$. The state of the population at the beginning of the step is $\bn$ and we assume that this state does not change until the end of the interval. We do however take into account the fact that the state of environment $\sigma$ can undergo changes in the interval from $t$ to $t+\Delta t$. As a consequence, $R_{r,\sigma}(\bn)$ (at fixed $\bn)$ is also a function of time.

We then introduce the time-averaged quantities
\be\label{eq:rbar}
\overline{R_r}(\bn) \equiv \frac{1}{\Delta t}\int_t^{t+\Delta t}\mathrm{d} t ~R_{r,\sigma(t)}(\bn),
\ee
noting that the time average is {\em over the duration of the time step only} as opposed to a long-term asymptotic time average. Given that the time step $\Delta t$ is finite ($\Delta t<\infty$) and assuming that the environment fluctuates with finite rates ($\lambda<\infty$), the quantity $\overline{R_r}(\bn)$ is a stochastic variable as it depends on the realisation of the environmental process. In one given time interval, the rates $\overline{R_r}(\bn)$ for different $r$ will be correlated as they all derive from the same path of the environment. As we will show below, the fluctuations of the random variables $\overline{R_r}(\bn)$ in any one time step are inversely proportional to $\lambda\Delta t$ to leading order. In the limit $\lambda \Delta t\to\infty$ the $\overline{R_r}(\bn)$ become deterministic.

We assume that the distribution for $\sigma$ at the beginning of the time step is the stationary distribution $\brho^*$. This is the case for example, if then environmental state is drawn from the stationary distribution at the beginning of the simulation. The distribution for $\sigma(t')$ is then also the stationary distribution at each time $t'\in[t,t+\Delta t]$. Writing $\avg{\dots}$ for the average over realisations of the environmental process, we then have
\be
\avg{\overline{R_r}(\bn)}=R_r^*(\bn),
\label{rates-avg-env}
\ee
with $R_r^*(\bn)$ as in Eq.~(\ref{eq:rstar}).

However, $\sigma(t')$ ($t'\in[t,t+\Delta t]$) will generally be correlated with $\sigma(t)$. Neglecting these means to operate in the adiabatic limit.  We would like to retain some of these correlations. In order to compute second moments  $\avg{\overline R_r(\bn) \overline R_s(\bn)}$ we first use Eq.~(\ref{eq:rbar}). This leads to averages of the type $\avg{R_{r,\sigma(t_1)}(\bn) R_{s,\sigma(t_2)}(\bn)}$, where $t_1$ and $t_2$ are two times in the interval from $t$ to $t+\Delta t$. The second moments can then be expressed in terms of $q_{\sigma\to\sigma'}(\cdot)$ as follows
\begin{align*}
&\avg{\overline R_r(\bn) \overline R_s(\bn)} = \\
&\frac{1}{\Delta t^2}\sum_{\sigma\sigma'}  \int_t^{t+\Delta t} \mathrm{d} t_1 \int_{t_1}^{t_1+\Delta t} \mathrm{d} t_2 \Big\{ \rho^*_\sigma q_{\sigma\to\sigma'}(t_2-t_1) \\
&\quad \quad \times \big[R_{r,\sigma}(\bn)R_{s,\sigma'}(\bn)+R_{r,\sigma'}(\bn)R_{s,\sigma}(\bn)\big]\Big\}. \numberthis
\label{avg-prod-Rs}
\end{align*}

Further details are given in Appendix \ref{app:second_moment}. Using Eq.~(\ref{eq:q}) we find
\begin{align*}
&\avg{\overline R_r(\bn) \overline R_s(\bn)} -R_r^*(\bn)R_s^*(\bn) = \\
& \frac{1}{\Delta t^2}\sum_{\sigma\sigma'}\sum_{\ell=2}^M \bigg\{\rho^*_\sigma c_{\ell,\sigma} v_{\ell,\sigma'} \\
& \times  \big[R_{r,\sigma}(\bn)R_{s,\sigma'}(\bn)+R_{r,\sigma'}(\bn)R_{s,\sigma}(\bn)\big] \\
&~~~~~~~~~~\times \int_t^{t+\Delta t}  \mathrm{d}t_1 \int_{t_1}^{t_1+\Delta t} \mathrm{d}t_2 ~e^{\lambda\mu_\ell (t_2-t_1)}\bigg\}.\numberthis
 \label{eq:rr0} 
 \end{align*}
For fixed $\ell\in\{2,\dots,M\}$ the integral in the last expression evaluates to
\begin{align*}
&\int_t^{t+\Delta t} \mathrm{d} t_1 \int_{t_1}^{t_1+\Delta t} \mathrm{d}t_2 ~e^{\lambda\mu_\ell (t_2-t_1)} = \\
&-\dfrac{\Delta t}{\lambda\mu_\ell}+\frac{1}{(\lambda\mu_\ell)^2}\left[e^{\lambda\mu_\ell \Delta t}-1\right].\numberthis
\end{align*}
For $\lambda\Delta t \gg 1$ the first term dominates after inserting into Eq.~(\ref{eq:rr0}), as also observed in \cite{hufton2016intrinsic}. We are then left with
\begin{align*}
&\avg{\overline R_r(\bn) \overline R_s(\bn)}-R_r^*(\bn)R_s^*(\bn) = \\
&-\frac{1}{\lambda\Delta t}\sum_{\sigma\sigma'}\sum_{\ell=2}^M  \bigg\{\frac{1}{\mu_\ell} R_{r,\sigma}(\bn)R_{s,\sigma'}(\bn)   \\
&~~~~ \times \big[\rho^*_\sigma c_{\ell,\sigma} v_{\ell,\sigma'}+\rho^*_{\sigma'} c_{\ell,\sigma'} v_{\ell,\sigma}\big] \bigg\} \label{eq:rr}. \numberthis
\end{align*}

The main challenge in implementing the $\tau$FE algorithm is then to find the average rates from Eq.~(\ref{rates-avg-env}) for all $r$, and the second moments from Eq.~(\ref{eq:rr}) for any pair $r, s$ of reactions affected by the environment. 

\subsection{Description of the algorithm}
\label{sec:algo-description}

Without loss of generality, we assume that only the rates for the reactions $r=1, \dots, L$ ($L \leq R$) depend on the environmental state $\sigma$. 

The $\tau$FE algorithm with time step $\Delta t$ proceeds as follows:
\begin{enumerate}
\item[1.] Initiate the population in state $\bn(0)$. Set time to $t=0$.
\item[2.] Compute $R^*_r(\bn)$ for $r=1, \dots, L$ using Eq.~(\ref{eq:rstar}), and the covariances $\Xi_{rs}(\bn)\equiv \avg{\overline R_r(\bn) \overline R_s(\bn)}-R_r^*(\bn)R_s^*(\bn)$ using Eq.~(\ref{eq:rr}) for every pair $r,s \in \{1, \dots, L \}$.
\item[3.] (i) First consider the reactions with rates dependent on the environment: Draw correlated Gaussian random numbers $\ell_1,\dots,\ell_L$ such that $\avg{\ell_r}=R^*_r(\bn)$, and $\avg{\ell_r \ell_s}-\avg{\ell_r}\avg{\ell_s}=\Xi_{rs}(\bn)$. If $\ell_r<0$ for any $r\in\{1,\dots,L\}$ set $\ell_r=0$. \\~\\
(ii) For the remaining reactions $r\in\{L+1,\dots, R\}$ set $\ell_r=R_r(\bn)$. These are the reactions with rates independent of the environment. 
\item[5.] Draw independent Poissonians random numbers $m_r$, $r=1,\dots, R$, each with parameter $\ell_r \Delta t$. 
\item[6.] Update the state of the population, $\bn(t+\Delta t)=\bn(t)+\sum_{r} m_r \bnu_r$. 
\item[7.] Increment time by $\Delta t$ and go to 2.
\end{enumerate}

We note that the mean of the $\ell_r$ in step 3(i) is of order $(\lambda \Delta t)^0$, and their variance of order $(\lambda \Delta t)^{-1}$. Truncation of the $\ell_r$ will therefore only be required very rarely when $\lambda\Delta t \gg 1$.

Evaluating the expressions in Eqs.~(\ref{eq:rstar}) and (\ref{eq:rr}) in step 2 requires eigenvalues $\mu_\ell$ of the transition matrix $\bA(\bn)$ for the environment, the eigenvectors, $\bv_\ell$ (including the stationary distribution $\bv_1=\brho^*$), and the coefficients $c_{\ell,\sigma}$ for all $\sigma$. If the environmental process is independent of the state of the population (the $A_{\sigma\to\sigma'}$ are not functions of $\bn$), then these quantities do not depend on $\bn$, and only need to be calculated once at the beginning.

In Sec.~\ref{sec:discrete} we first test the $\tau$FE algorithm on different models with discrete environments. However that the algorithm can also be extended to the case of environmental dynamics with continuous states. This will be discussed in Sec.~\ref{sec:continuous}.

\section{Application of the $\tau$FE algorithm to models with discrete environmental states} \label{sec:discrete}
We now consider three examples of systems with discrete environmental states. 

The first example (Sec.~\ref{sec:ex1}) is a genetic circuit. The role of the environment is here played by a process of binding and unbinding to promoters of the genes described by the model. Gene regulatory systems can exhibit time scale separation as discussed for example in \cite{Gunawardena, BuchlerGerlandHwa, Lin2018}. Mathematically the model describes a population with two types of individuals and an environment with two states (bound/unbound). The environmental dynamics depends on the state of the population. 

The second example (Sec.~\ref{sec:ex2}) is a toy model with two species in the population and three environmental states. The environmental process in this example is independent of the state of the population. 

Sec.~\ref{sec:ex3} finally focuses on a bimodal genetic switch with two species in the population, and an environmental process with three states, and with rates which depend on the state of the population.

\subsection{Genetic circuit: two system-independent environments, two species}\label{sec:ex1}

This system models the dynamics of two genes, which produce two different regulatory proteins: X (a transcription factor) and Y (an inhibitor that titrates $X$ into an inactive complex). Specifically, we use the activator-titration circuit described in \cite{lin2018efficient}. The reactions are as follows:
\begin{align*}
&\emptyset  \xrightarrow{\Omega \beta_X} X, \quad \emptyset  \xrightarrow{\Omega \beta_{Y,\sigma} } Y, \quad (\sigma = 0,1) \\
&X \xrightarrow{\delta_X} \emptyset, \quad Y \xrightarrow{\delta_Y} \emptyset, \\
&X + Y \xrightarrow{\alpha/\Omega} \emptyset,\\
&E_0 \xrightarrow{\lambda n_X \kappa_Y/\Omega} E_1, \quad \text{and,} \quad E_1 \xrightarrow{\lambda \theta_Y} E_0,\numberthis
\label{gen-circuit-processes}
\end{align*}
where the $E_{\sigma}$ denote states of the environment $(\sigma = 0,1)$. These two environmental states represent situations in which a transcription factor $X$ is bound to the promoter of gene $Y$ (state $E_1$), or no transcription factor is bound ($E_0$), respectively. The first two reactions in Eq.~(\ref{gen-circuit-processes}) describe the production of the two proteins ($X$ and $Y$). The production rates are $\beta_X$ and $\beta_{Y,\sigma}$. The former is independent of the environmental state, the latter explicitly depends on $\sigma$ (i.e., on the presence or absence of a bound transcription factor). The reactions in the second line of Eq.~(\ref{gen-circuit-processes}) describe degradation of $X$ and $Y$, and the reaction in the third line captures titration. The binding and unbinding processes of the transcription factor are described by the reactions in the last line. The parameter $\Omega$ in the reaction rates determines the typical number of particles in the system, for further details see \cite{lin2018efficient}. We write $n_X$ for the number of $X$-particles in the system, and similarly $n_Y$ is the number of $Y$-particles. One finds $n_X, n_Y={\cal O}(\Omega)$ in the stationary state.

 \begin{figure}[t]
  \centering
    \includegraphics[width=0.475\textwidth]{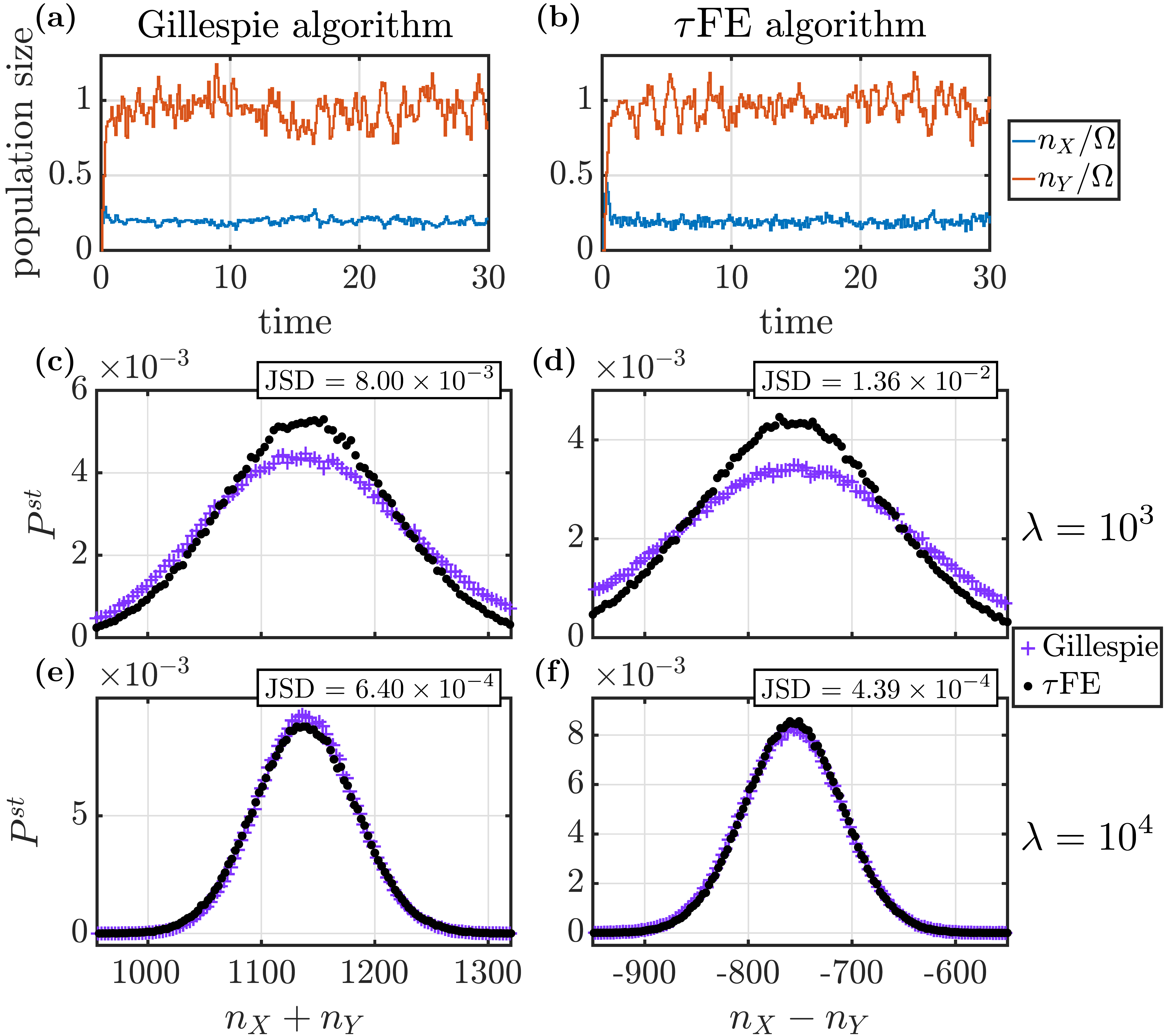}
    \caption{Simulation output for the model of the genetic-circuit in Eq.~(\ref{gen-circuit-processes}). Panel (a) shows a sample path obtained from Gillespie simulations of the full model. Panel (b) is a sample path from the $\tau$FE algorithm [$\lambda = 10^3$ in panels (a) and (b)]. Panels (c) and (d) show the stationary distributions of $n_X + n_Y$ and $n_X - n_Y$, respectively, for $\lambda=10^3$, while (e) and (f) are for $\lambda = 10^4$. In each panel (c)--(f) we report the Jensen-Shannon divergence (JSD) between the distributions obtained using the two different simulation methods. Remaining parameters: $\Omega = 10^3, \beta_X = 2, \beta_{Y,0} = 0, \beta_{Y,1} = 10, \delta_X  = \delta_Y = 1, \kappa_Y = 1, \theta_Y = 0.5, \text{and } \alpha = 10$. For the $\tau$FE we have used a time step $\Delta t = 0.1$.}
    \label{genetic-circuit}
\end{figure}

Mathematically, the model consists of two species in the population (with numbers of particles $n_X, n_Y$), and two environmental states, $\sigma=0,1$.  We therefore have $S=2,M=2$. Eqs.~(\ref{rates-avg-env}) and (\ref{eq:rr}) can be evaluated explicitly for this case, see also \cite{hufton2019classical}. The only process affected by the state of the environment is the production of $Y$, with rate $\beta_{Y,\sigma}$. This rate becomes a (clipped) Gaussian random variable in the $\tau$FE algorithm, with first moment 
\begin{equation}
 \avg{\overline{\beta}_Y}=\beta_Y^*= \dfrac{\theta_Y  \beta_{Y,0}  + n_X \kappa_Y \beta_{Y,1}/\Omega}{\theta_Y + n_X \kappa_Y/\Omega},
\end{equation}
and with variance
\begin{align*}
\sigma_{\beta_Y \beta_Y}^2 &\equiv \langle \bar{\beta}_Y^2 \rangle - {\beta_{Y}^*}^2 \\
&= \dfrac{2 n_X  \kappa_Y \theta_Y/\Omega}{\lambda \Delta t (n_X \kappa_Y/\Omega + \theta_Y)} \left(\beta_{Y,0} - \beta_{Y,1} \right)^2. \numberthis
\end{align*}
Further details of the derivation can be found in Appendix~\ref{app:two-sp-two-env}.

Simulation results for this model are shown in Fig.~\ref{genetic-circuit}. In panels (a) and (b) we illustrate typical sample paths obtained from Gillespie simulations of the full model (population and environment), and from the $\tau$FE algorithm, respectively. We also show the stationary distributions for the quantities $n_X + n_Y$ and $n_X - n_Y$ as obtained from both simulation algorithms. The distributions in panels (c) and (d) are for $\lambda=10^3$ (i.e., moderately fast environmental dynamics), there are then remaining discrepancies between the $\tau$FE algorithm and simulations of the full model. In panels (e) and (f) the time-scale separation is larger ($\lambda=10^4$). The agreement improves as indicated by the Jensen-Shannon divergence (JSD) \cite{JSD1,JSD2} given in the figure. 

We note at this point that the average CPU time to run a sample path up to time $t=10^3$ with parameters as in Fig.~\ref{genetic-circuit} (e) and (f) is $2.94$ seconds for the Gillespie algorithm, and $0.03$ seconds for the $\tau$FE algorithm (with a time step $\Delta t=0.1$). These average simulation times are obtained from ten runs. They indicate that the $\tau$FE algorithm can significantly increase efficiency while producing results of the quality shown in Fig.~\ref{genetic-circuit}. We stress that our primary interest is the relative comparison of computing times, and not on absolute simulation times \footnote{For completeness, we add that simulations were performed on a MacBook Pro (Mid 2014), with processor 2.6 GHz Dual-Core Inter Core i5, and memory 8 GB 1600 MHz DDR3.}.

\subsection{Birth-death process: three environments, two species}\label{sec:ex2}

Next, we consider a two-species birth-death process subject to an external environment which can be in one of three different states. This is a toy model chosen for illustration and does not represent any specific natural system. However, it captures elements of models of population dynamics.

The species in the population are labeled $A$ and $B$, and the environmental states $\sigma=0,1,2$. Particles of type $A$ are produced with rate $\Omega \alpha_{\sigma}$, and particles of type $B$ with rate $\Omega \beta_{\sigma}$. The subscript $\sigma$ indicates explicit dependence on the state of the environment. Particles are removed with constant per capita rates $d_A$ and $d_B$ respectively. The parameter $\Omega$ again sets the typical size of the population. We write $n_A$ and $n_B$ for the number of individuals of either species. The environmental states cycle stochastically through the sequence $\sigma = 0, 1, 2, 0, \dots$, with rate constants $\lambda k_1$, $\lambda k_2$ and $\lambda k_0$ for the three transitions. Mathematically, the reactions in this model are
\begin{align*}
& \emptyset \xrightarrow{\Omega \alpha_{\sigma}} A, \quad \emptyset \xrightarrow{\Omega \beta_{\sigma}} B, \quad (\sigma=0,1,2)\\
& A \xrightarrow{d_A} \emptyset, \quad B \xrightarrow{d_B} \emptyset,\\
& E_0 \xrightarrow{\lambda k_1} E_1, \quad E_1 \xrightarrow{\lambda k_2} E_2, \quad  E_2 \xrightarrow{\lambda k_0} E_0, \numberthis
\label{three-env-two-sp-procs}
\end{align*}
where as before $E_{\sigma}$ denotes the environment. The rates $k_0, k_1,$ and $k_2$ are constant parameters, independent of the population state.

Details of the calculation of the average rates and their second moments can be found in Appendix~\ref{app:three-env-two-sp}. The average production rates for the two types of particles are
\BE
\alpha^*&=& \dfrac{k_0 \alpha_0 + k_1 \alpha_1 + k_2 \alpha_2}{k_0 + k_1 + k_2}, \nonumber \\
\beta^* &=& \dfrac{k_0 \beta_0 + k_1 \beta_1 + k_2 \beta_2}{k_0 + k_1 + k_2},
\label{alpha-beta-star}
\EE
while the covariance $\sigma_{\alpha \beta} = \sigma_{\beta \alpha} \equiv \langle \bar{\alpha} \bar{\beta} \rangle - \alpha^* \beta^*$ takes the form
\begin{align*}
\sigma_{\alpha \beta} &= \dfrac{\theta^2}{\lambda \Delta t}  \Big \{ (\alpha_0 - \alpha_1)(\beta_0 - \beta_1) \left(3 k_0^2 - k_{0,1} k_{0,2}\right) \\
& \quad + (\alpha_1 - \alpha_2)(\beta_1 - \beta_2) \left(3 k_1^2 - k_{1,0} k_{1,2}\right) \\
& \quad +(\alpha_0 - \alpha_2)(\beta_0 - \beta_2) \left(3 k_2^2 - k_{2,0} k_{2,1}\right) \Big \}, \numberthis
\label{cov-three-env}
\end{align*}
with
\begin{equation}
\theta^2 \equiv \dfrac{k_0 k_1 k_2}{(k_0 k_1 + k_1 k_2 + k_2 k_0)^3},
\end{equation}
and $k_{\sigma, \sigma'} = k_{\sigma} - k_{\sigma'}$, for $\sigma, \sigma' \in \{0,1,2\}$. The variance $\sigma_{\alpha \alpha} \equiv \langle \bar{\alpha} \bar{\alpha} \rangle - (\alpha^*)^2$, is obtained by replacing all instances of $\beta_\sigma$ on the right-hand side of Eq.~(\ref{cov-three-env}) with $\alpha_\sigma$. The analog $\sigma_{\beta\beta}$ is obtained similarly by replacing $\alpha_\sigma$ with $\beta_\sigma$. 

\begin{figure}[t]
  \centering
      \includegraphics[width=0.475\textwidth]{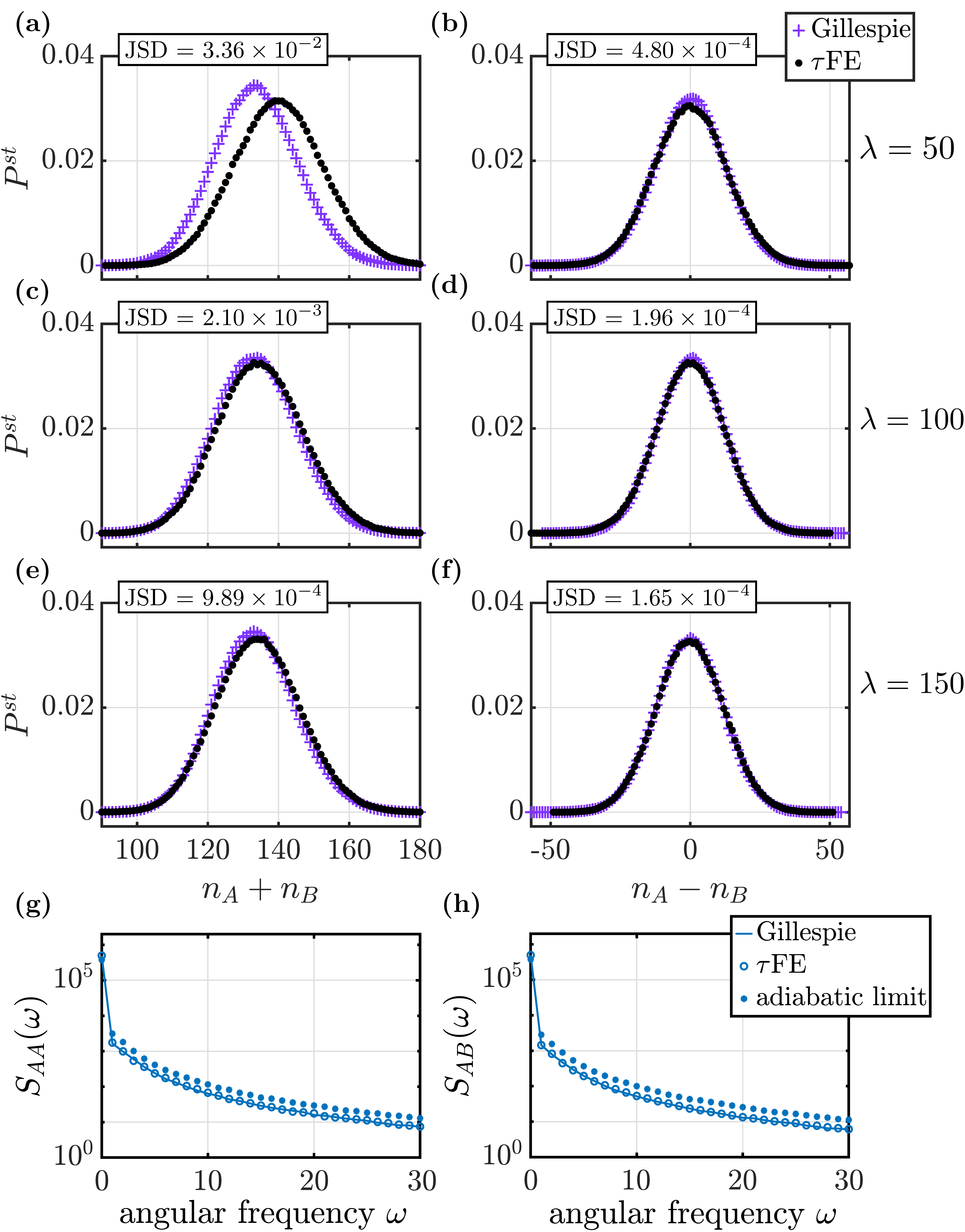}
    \caption{Simulation output of the two-species birth-death process in an environment with three states [Eq.~(\ref{three-env-two-sp-procs})]. Panels (a), (c), and (e) show the stationary distribution of $n_A + n_B$ obtained using the Gillespie algorithm for the full model, and the  $\tau$FE algorithm. Data is shown for different values of $\lambda$. Panels (b), (d), and (f) show the stationary distribution of $n_A - n_B$. Parameters used: $k_0 = k_1 = k_2 = 1, \Omega = 20, d_A = d_B = 0.1$, $\alpha_0 = \beta_0 = \beta_1 = \alpha_2 = 0$, and $\alpha_1 = \beta_2 = 1$.  In each panel (a)--(f) we report the Jensen-Shannon divergence (JSD) between the two distributions obtained from Gillespie simulations of the full model and from the $\tau$FE algorithm. Panels (g) and (h): Spectral densities $S_{AA}(\omega)$ and $S_{AB}(\omega)$ [Eq.~(\ref{spectral-def})] obtained from simulations using the Gillespie algorithm (full line), the $\tau$FE algorithm (open circles), and from conventional $\tau$-leaping simulations of the model in the adiabatic limit (asterisks). Parameters in (g) and (h) are as in panels (e) and (f), i.e., $\lambda = 150$. The time step for the $\tau$FE algorithm and for conventional $\tau$-leaping in the adiabatic limit is $\Delta t=0.1$.}
    \label{pst-two-sp-three-env}
\end{figure}

In Fig.~\ref{pst-two-sp-three-env} we report results from numerical simulations for this model, both from conventional Gillespie algorithm of the full systems of environment and population, and using the $\tau$FE algorithm. Panels (a)--(f) show the stationary distributions of $n_A + n_B$ and $n_A - n_B$. As seen from the data for example in panel (a) the $\tau$FE algorithm displays deviations from Gillespie simulations of the full model when the environmental process is not sufficiently fast. We quantify these deviations again through the  Jensen--Shannon divergence between the two distributions. The deviations reduce as the time scale separation $\lambda$ is increased, i.e., when the environmental process becomes faster relative to the dynamics within the population.

 \begin{figure*}[t]
  \centering
    \includegraphics[width=0.9\textwidth]{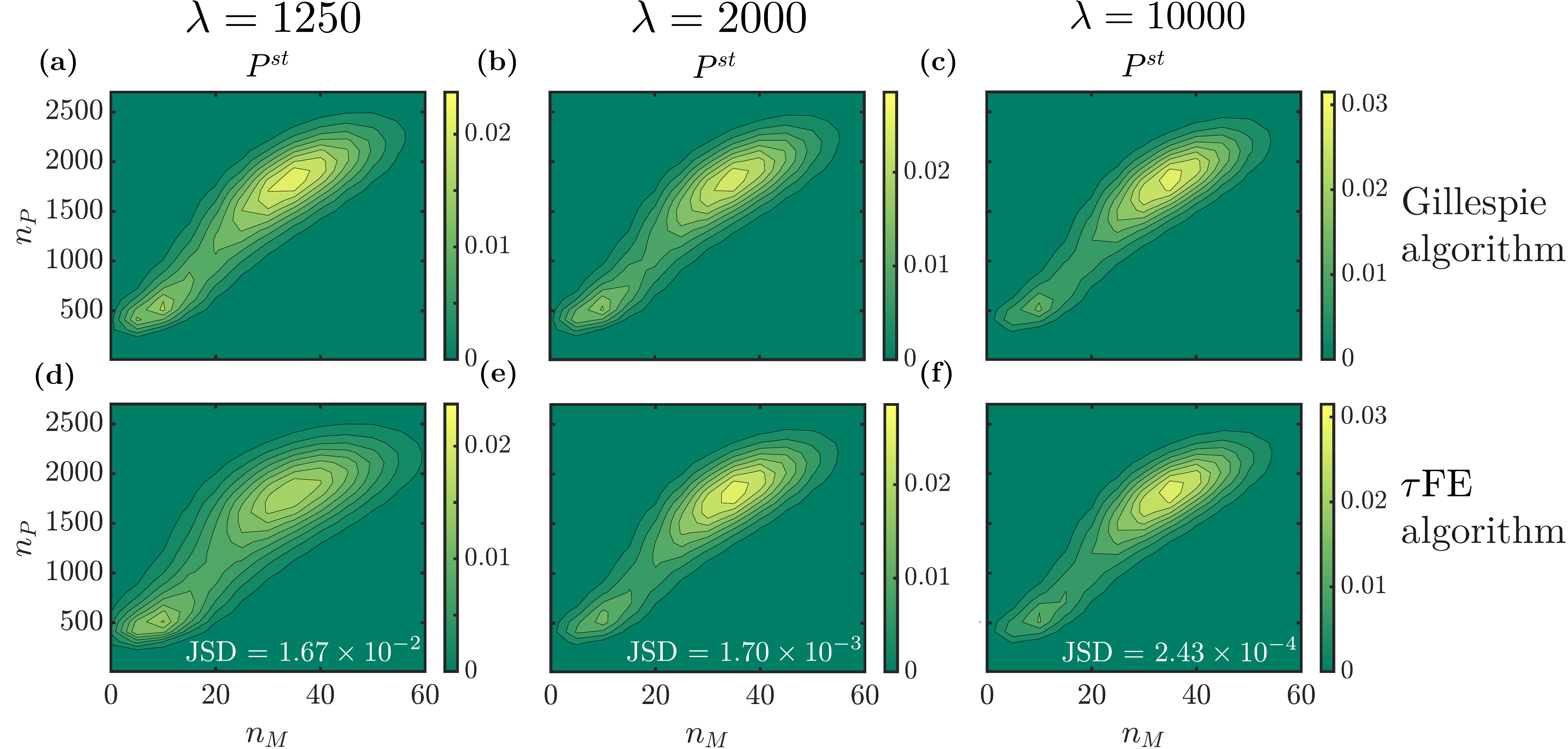}
    \caption{Stationary distribution for the numbers of mRNA and protein molecules ($n_M$ and $n_P$, respectively) in the model of a genetic switch [Eq.~\ref{three-env-two-sp-procs}]. Data is shown for different values of $\lambda$ and for simulations of the full model by means of the Gillespie algorithm, and the $\tau$FE algorithm. Parameters: $N=2, \Omega = 50, b_0 = b_1 = 1, b_2 = 20, d = 9.2, \beta  = 50, \delta = 1, k_+ = 0.025, k_- = 1$. The stationary distribution is obtained from a long run up to time $t=10^5$. For the $\tau$FE algorithm we use $\Delta t = 100/\lambda$. For each of the three values of $\lambda$ we report the Jensen-Shannon divergence (JSD) between the distributions obtained from the two simulation methods.}
    \label{bim-gen-sim-both}
\end{figure*}

In order to examine if the $\tau$FE algorithm accurately reproduces dynamical features (i.e., properties of the system beyond the stationary distribution), we show spectral densities of the time series for $n_A$ and $n_B$ in Fig.~\ref{pst-two-sp-three-env}(g) and (h). The spectral densities are defined as
\begin{align*}
&S_{AA}(\omega) = \langle |\hat{n}_A(\omega)|^2 \rangle, \\
& S_{AB}(\omega) = \langle \hat{n}_A^{\dagger}(\omega) \hat{n}_B(\omega) \rangle, \numberthis
\label{spectral-def}
\end{align*}
where $\hat{n}_A(\omega)$ and $\hat{n}_B(\omega)$ are the Fourier transforms of $n_A(t)$ and $n_B(t)$, respectively. The dagger denotes complex conjugation. The data from the $\tau$FE algorithm (open symbols) in Fig.~\ref{pst-two-sp-three-env}(g) and (h) compares well with spectra obtained from direct Gillespie simulations of the full model (solid lines). This shows the $\tau$FE method indeed captures the dynamics of $n_A$ and $n_B$. We also provide a comparison against the spectral densities obtained from conventional $\tau-$leaping simulations in the adiabatic limit, i.e., simulations with constant rates $\alpha^*$ and $\beta^*$ for the production events [Eq.~(\ref{alpha-beta-star})]. These are shown as full markers in Fig.~\ref{pst-two-sp-three-env}(g) and (h). One then finds more substantial systematic deviations. This is because environmental fluctuations are discarded in the adiabatic limit. The $\tau$FE algorithm on the other hand captures the stochasticity of the environment to sub-leading order in $\lambda^{-1}$ in each iteration step.

\subsection{Bimodal genetic switch: three system-state dependent environments, two species}\label{sec:ex3}

We now consider a model studied in \cite{LinHufton2018,hufton2019model}, describing a single gene $G$ with a promoter site which can bind to a total of up to $N$ molecules of protein. The number of protein molecules bound, $\sigma$, plays the role of the environment in this setting. The rate for transitions from $\sigma$ to $\sigma + 1$ depends on the number of protein molecules. The reactions in this model can be summarised as follows,
\begin{align*}
G_\sigma  + P \xrightleftharpoons[\lambda k_-]{\lambda k_+/\Omega} G_{\sigma + 1},& \quad \text{for} \quad \sigma < N, \\
G_\sigma \xrightarrow{\Omega b_\sigma} G_{\sigma} + M, & \\
M  \xrightarrow{d} \emptyset, \quad M  \xrightarrow{\beta} M + P,&  \quad P \xrightarrow{\delta} \emptyset, \numberthis
\label{bimodgen-reac}
\end{align*}
where $M$ and $P$ refer to molecules of mRNA and protein, respectively.  The production rate $b_\sigma$ for mRNA depends on the number of protein molecules bound to the promoter. We refer to  \cite{LinHufton2018,hufton2019model} for further details. In the following we write $n_M$ and $n_P$ for the numbers of particles of either type. One interesting feature of this model is that the distribution of the protein and mRNA populations can become bimodal, as illustrated in Fig.~\ref{bim-gen-sim-both}. This leads to bistability, with trajectories transitioning between the two modes of the joint distribution of $n_P$ and $n_M$.  Hence, the model describes a genetic switch.

In this model only the production rate of mRNA molecules is affected by the state of the environment. The average mRNA-production rate is found as
\begin{equation}
b^* = \dfrac{k_-^2 b_0 + k_- \tilde{k}_+ b_1 + \tilde{k}_+^2 b_2}{k_-^2 + k_- \tilde{k}_+ +  \tilde{k}_+^2},
\label{b-star-bim-gen-swt}
\end{equation}
with $\tilde{k}_+ = k_+ n_P/\Omega$. The second moment of the production rate takes the form
\begin{align*}
 \sigma_{b b}^2 & \equiv \langle \bar{b}^2  \rangle - {b^*}^2 \\
&=\dfrac{\theta^2}{\lambda \Delta t} \Big \{(b_0 - b_1)^2 k_- \left(k_-^2 +  \tilde{k}_+ k_- - \tilde{k}_+^2 \right)\\
& \quad +(b_0 - b_2)^2 2 k_- \tilde{k}_+ \left( k_- + \tilde{k}_+ \right)  \\
& \quad +(b_1 - b_2)^2 \tilde{k}_+ \left(\tilde{k}_+^2 + \tilde{k}_+ k_- -  k_-^2 \right)\Big \}, \numberthis
\label{cov-bim-gen-swt}
\end{align*}
with 
\be
\theta^2 = \dfrac{2 k_- \tilde{k}_+}{\left(k_-^2 + k_- \tilde{k}_+ + \tilde{k}_+^2 \right)^3}.
\ee
Details of the calculation leading to Eqs.~(\ref{b-star-bim-gen-swt}) and (\ref{cov-bim-gen-swt}) can be found in Appendix~\ref{app:bim-gen-swt}.

Figure~\ref{bim-gen-sim-both} shows the stationary joint distribution of the number of mRNA and protein molecules for different values of the time-scale separation parameter $\lambda$. The figure shows data from Gillespie simulations of the full model [panels (a)--(c)], and data from the $\tau$FE algorithm [panels (d)--(f)]. The $\tau$FE algorithm captures the distribution profile with two local maxima. For low values of $\lambda$ (i.e., a relatively slow environmental process) the distribution obtained from $\tau$FE tends to be wider than those from the Gillespie algorithm. The agreement improves for faster environments, as indicated again by the  Jensen--Shannon distances in Fig.~\ref{bim-gen-sim-both}. 

 \begin{figure}[t]
  \centering
    \includegraphics[width=0.475\textwidth]{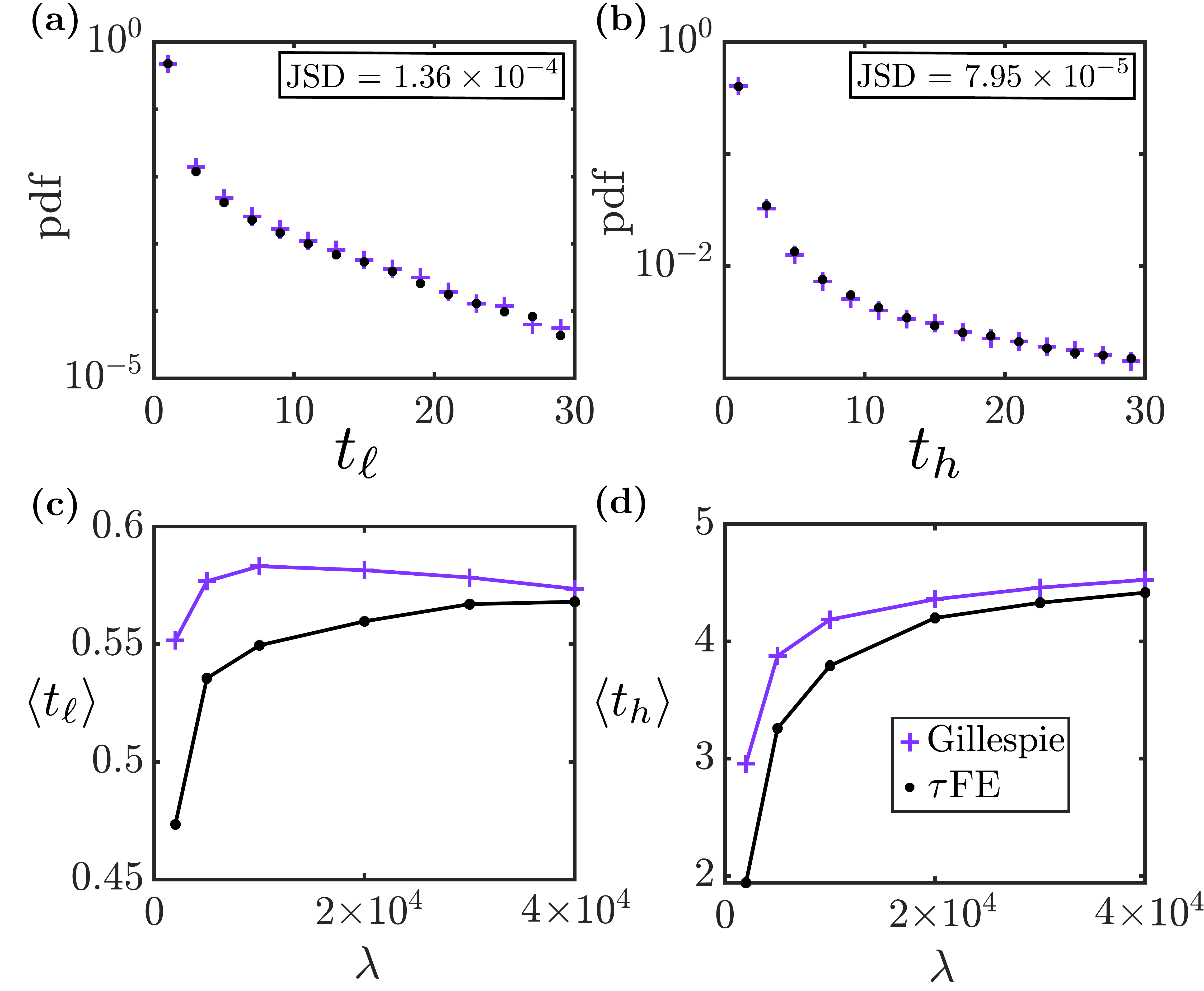}
    \caption{Sojourn times $t_\ell$ and $t_{\rm h}$ near the two modes of the bistable genetic switch (see also Fig.~\ref{bim-gen-sim-both}). Panels (a) and (b) show the distribution of the time spent in the vicinity of each mode (see text for details); data obtained from $\tau$FE algorithm is shown along with results from exact Gillespie simulations of the full model ($\lambda=2000$). In each panel we report the Jensen--Shannon distance between the two distributions. Panels (c) and (d) show the mean sojourn times as a function of the time-scale parameter $\lambda$. The parameters are as in Fig.~\ref{bim-gen-sim-both}, the lower mode is $(n_M, n_P) = (10, 500)$, and the upper mode $(n_M, n_P) = (30, 1800)$. For the $\tau$FE algorithm we use a time step of $\Delta t = 100/\lambda$.}
    \label{waiting-times-bim-gen-swt}
\end{figure}

In Fig.~\ref{waiting-times-bim-gen-swt}, we show the distribution and means of the sojourn times $t_{\ell}$ and $t_{\rm h}$ near the lower and higher modes of the stationary distribution. More precisely this is the time between entering and leaving a designated region around each of the modes. The lower maximum of the stationary distribution is sharper than the upper maximum (Fig.~\ref{bim-gen-sim-both}). Accordingly, we have chosen a smaller region at the lower mode than at the upper mode. For the lower mode, we use the region $0 \leq n_M \leq 20$, $0 \leq n_P \leq 1100 $ which encloses the mode at $(n_M, n_P) = (10,500)$. For the higher mode we use the region $20 \leq n_M \leq 80$, $1100 \leq n_P \leq 2700$ enclosing the mode at $(n_M, n_P) = (30,1800)$. 

The data shown in the figure is constructed from one long sample path (run until $t=10^6$), recording the points in time at which the system enters or leaves either region. Gillespie simulations operate in continuous time and the $\tau$FE algorithm in discrete time. In order to remove any artefacts resulting from this difference, the same time resolution ($0.05$) is used in both algorithms for the measurement of arrival and departure times. Because the lower mode is sharper than the upper maximum and because the sizes of the two detection regions are different the sojourn time $t_{\ell}$ at the lower mode is found to be smaller than that at the higher mode, $t_{\rm h}$.

The distributions of sojourn times in Figs.~\ref{waiting-times-bim-gen-swt} (a) and (b) indicate that the $\tau$FE algorithm captures this dynamic quantity, provided the environmental process is sufficiently fast. This is confirmed in panels (c) and (d), where we show the mean sojourn times as a function of the relative speed $\lambda$ of the environment compared to the population dynamics.  As seen in both panels, the $\tau$FE algorithm generates accurate measurements of the mean sojourn times $\avg{t_{\ell}}$ and $\avg{t_{\rm h}}$ in the limit $\lambda\gg 1$. 

At the same time, stochastic effects due to the random environmental process are captured for large but finite $\lambda$. This can be seen in Fig.~\ref{waiting-times-bim-gen-swt} (d): the mean sojourn time $\avg{t_{\rm h}}$ drops significantly as the environmental process becomes slower, and hence additional noise is injected into the population (there is no environmental noise in the adiabatic limit). While there are quantitative differences compared to exact simulations, the $\tau$FE algorithm captures this reduction of $\avg{t_h}$. Panel (c) reveals that there are also limitations to the precision of the $\tau$FE algorithm. The mean sojourn time $\avg{t_{\ell}}$ near the lower mode is affected much less by a reduction of the time-scale separation parameter $\lambda$ than the mean sojourn time at the upper mode. This indicates that the escape from this region is driven mostly by intrinsic noise rather than by environmental stochasticity. While the data from the two algorithms remains within approximately $10\%$ for sufficiently fast environmental dynamics ($\lambda\gtrsim 10^4$) the $\tau$FE algorithm is unable to capture the small rise of $\avg{t_{\ell}}$ observed in Gillespie simulations for intermediate values of $\lambda$.

\begin{table}[b]
\centering
\begin{tabular}{ m{6em} m{6em}  m{6em}  } 
\hline
\hline
$\lambda$ & Gillespie  & $\tau$FE  \\ 
\hline
1250& 1.35 & 0.04\\ 
\hline
2500 & 1.89 & 0.08\\ 
\hline
5000 & 2.62 & 0.17\\ 
\hline
10000 & 4.30 & 0.31 \\
\hline
20000 & 7.77 & 0.57 \\
\hline
\hline
\end{tabular}
\caption{Mean computation time (in seconds) required to simulate one sample path up to $t=10^3$ of the bimodal genetic-switch system defined in Eq.~(\ref{bimodgen-reac}). Measurements are from ten independent sample runs, using Gillespie simulations of the full model, and the $\tau$FE algorithm respectively. Parameters are as in Fig.~\ref{bim-gen-sim-both}. For the $\tau$FE algorithm we set $\Delta t = 100/\lambda$. }
\label{table-bim-gen-swt}
\end{table}

In Table~\ref{table-bim-gen-swt} we compare the the computing time required for both the Gillespie algorithm and the $\tau$FE method for different values of $\lambda$. The data in the table is the CPU time required to generate one sample path up to time $t=10^3$, averaged over ten runs. The model parameters are as in Figs.~\ref{bim-gen-sim-both} and \ref{waiting-times-bim-gen-swt}. 

The full model comprises the reactions in the population and the environmental switching. The rates for the former reactions are independent of $\lambda$, the rates for the latter scale linearly in $\lambda$. Accordingly, one expects the computing time for Gillespie simulations of the full model to be linear in $\lambda$, with a non-zero intercept. The data in the table is consistent with this. We note that Gillespie algorithm does not require any time discretisation.

The running time for the $\tau$FE algorithm depends on the choice of the time step. The time step in turn affects the accuracy of the outcome. If $\Delta t$ is large, then $\tau$FE simulations are fast, but the approximation to the continuous-time full model becomes less good. On the other hand the time step must not be too small, as the construction of the algorithm requires sufficient averaging of the environmental process in each step [Eqs.~(\ref{eq:rr0})--(\ref{eq:rr})]. The time step for the $\tau$FE algorithm in  Figs.~\ref{bim-gen-sim-both} and \ref{waiting-times-bim-gen-swt}, and in Table~\ref{table-bim-gen-swt} is chosen inversely proportional to $\lambda$. This is to ensure that each time step captures a sufficient number of switches of the environmental state. Accordingly, we expect the computing time for the $\tau$FE algorithm to scale linearly in $\lambda$, with no intercept. Again, the running times we measured in our simulations are consistent with this expectation. Overall, Table ~\ref{table-bim-gen-swt} shows that the $\tau$FE algorithm is able to generate data of the accuracy as in Figs.~\ref{bim-gen-sim-both} and \ref{waiting-times-bim-gen-swt} while reducing the computing effort approximately ten fold compared to full Gillespie simulations.

\section{Numerical simulation of continuous-environmental systems} \label{sec:continuous}
\subsection{Setup}
We turn now to systems which are subject to an environment with continuous states. Specifically, we follow \cite{assaf2013extrinsic} and assume that the environmental state $\sigma$ follows an Ornstein--Uhlenbeck process (see also \cite{roberts2015dynamics, assaf2}),
\begin{equation}
\frac{\mathrm{d} \sigma}{\mathrm{d} t} =  \lambda (m - \sigma) + \sqrt{2 \lambda v^2} \, \eta(t),
\label{ou-langevin}
\end{equation}
where $\eta(t)$ is Gaussian white noise of unit amplitude, in particular $\avg{\eta(t)\eta(t')}=\delta(t-t')$. The parameter $m$ is the average value of $\sigma$ in the long run, whilst $v$ controls the magnitude of noise. As before, the parameter $\lambda>0$ indicates how quickly the environment changes relative to the dynamics in the population; $\lambda$ is the equivalent of $1/\tau_c$ in the notation of \cite{assaf2013extrinsic}.

The probability distribution of finding the environment in state $\sigma$ at time $t$, given that was in state $\sigma'$ at time $t'$, can be obtained from the Fokker--Planck equation for the Ornstein--Uhlenbeck process, and is given by (see e.g. \cite{klebaner2005introduction, risken1996fokker})
\begin{align*}
&q_{\sigma'\to\sigma}(t-t')= \sqrt{\dfrac{1}{2 \pi v^2 (1-e^{-2\lambda(t-t')})}}\\
& \times \exp \left[- \frac{\left(\sigma - \sigma' e^{- \lambda(t-t')} - m \left(1 - e^{-\lambda(t-t')}\right)\right)^2}{2 v^2 \left(1 - e^{-2\lambda(t-t')} \right)}\right].      
        \numberthis \label{eq:oukernel}
\end{align*}

For $t\to\infty$ (and $t'$ fixed) this quantity tends to the stationary distribution
\begin{equation}
\rho^*_{\sigma} = \sqrt{\dfrac{1}{2 \pi v^2}} \exp \left[- \dfrac{\left(\sigma - m \right)^2}{2 v^2} \right].
\label{pst-sigma-eq}
\end{equation}
We note that it is not a requirement for the $\tau$FE algorithm that the environment follows an Ornstein--Uhlenbeck process. However, both functions $q_{\sigma'\to\sigma}(t-t')$ and $\rho^*_{\sigma}$ are required, as discussed in more detail below.

We proceed to describe how the $\tau$FE algorithm can be implemented for models with continuous environments (Sec.~\ref{sec:implem-algo}). 

In the case of discrete environments, continuous-time sample paths of the full model can be generated using the conventional Gillespie algorithm. This is an exact procedure: the ensemble of these sample paths faithfully describes the statistics of the full model. In Sec.~\ref{sec:discrete} we have used this as a benchmark to test the $\tau$FE algorithm. We are not aware of any analogous exact simulation method for models of discrete populations in a stochastic environment with continuous states. In order to test the $\tau$FE algorithm we therefore compare outcomes against those from approximation methods to generate paths of the combined set of the population and the environment. Several such methods exist, we describe these in Sec.~\ref{sec:contsim}. The tests of the $\tau$FE algorithm against the baseline of these methods are described in Sec.~\ref{sec:contsim_examples}.

\subsection{Implementation of the $\tau$FE algorithm for continuous environments}
\label{sec:implem-algo}

We proceed similar to discrete case in Sec.~\ref{sec:algo}, replacing the sums over $\sigma$ in Eqs.~(\ref{eq:rstar}) and (\ref{avg-prod-Rs}) with integrals. We then have
\be
R_r^*(\bn)
=\int_{-\infty}^{\infty} \mathrm{d} \sigma \rho^*_\sigma R_{r,\sigma}(\bn),
\label{FSA-cont-avg}
\ee
and the relation for the second moments turns into
\begin{align*}
&\avg{\overline R_r(\bn) \overline R_s(\bn)} = \\
&\frac{1}{\Delta t^2} \int_{-\infty}^{\infty} \mathrm{d} \sigma \int_{-\infty}^{\infty} \mathrm{d} \sigma'  \int_t^{t+\Delta t} \mathrm{d}t_1 \int_{t_1}^{t+\Delta t} \mathrm{d}  t_2 \\
&\times \Big\{\rho^*_\sigma q_{\sigma\to\sigma'}(t_2-t_1) \\
& \big[R_{r,\sigma}(\bn)R_{s,\sigma'}(\bn) +R_{r,\sigma'}(\bn)R_{s,\sigma}(\bn)\big]\Big\}. \numberthis
\label{FSA-cont-second-mom}
\end{align*}

Depending on the form of the stationary distribution $\rho^*_\sigma$, the kernel $q_{\sigma\to\sigma'}(t_2-t_1)$ and the rates $R_{r,\sigma}(\bn)$ the integrals in Eqs.~(\ref{FSA-cont-avg}) and (\ref{FSA-cont-second-mom}) can be carried out, and closed-form analytical expressions can be obtained. In Sec.~\ref{sec:contsim_examples} we explore a number of different examples, further scenarios are also discussed Appendix \ref{app:f}. Once the average rates and the second moments are calculated, the $\tau$FE algorithm is implemented as described in Sec.~\ref{sec:algo-description}.

\subsection{Conventional simulation approaches for discrete populations in continuous environments}\label{sec:contsim}
In this section we summarise `conventional' approaches to simulating discrete Markovian systems subject to environmental dynamics with continuous states. By `conventional' we mean methods which produce explicit (approximate) sample paths of the environmental process. This is in contrast to the $\tau$FE algorithm, which generates paths only of the system proper.

\subsubsection{Gillespie algorithm with discretised environmental states (GADE)}\label{sec:gillespie_discrete}

This approach is based on a discretisation of the space of environmental states, time remains continuous. Once such a discretisation for the environmental states is carried out, the combined states of the population and environment are also discrete. Simulations can be carried out using the conventional Gillespie method.  We will refer to this method as GADE (Gillespie approach with discretised environment). 

The key step in this approach is to find an appropriate dynamics in the space of discretised environmental states. We describe this in the context of the Ornstein--Uhlenbeck process in Eq.~(\ref{ou-langevin}). We discretise the environmental state into integer multiples of $\Delta\sigma$, i.e., the environment takes states $\dots,-2\Delta\sigma,-\Delta\sigma,0,\Delta\sigma,2\Delta\sigma,\dots$. Transitions from one state $k\Delta\sigma$ can only occur to states $(k\pm1)\Delta\sigma$. The transition rates are constructed such that this discrete process recovers the continuous Ornstein--Uhlenbeck dynamics in the limit $\Delta\sigma\to 0$.  The details of the construction are described in Appendix \ref{app:gillespie-cont}, we here only report the main outcome. Specifically, the rates to transition from state $k\Delta \sigma$ to $(k\pm 1)\Delta \sigma$ can be chosen as
\be
T_k^\pm=\frac{\lambda}{2\Delta\sigma}\left[\pm (m-k\Delta\sigma)+\frac{2 v^2}{\Delta\sigma}\right].
\label{discrete-sigma-rates}
\ee
This process can then be simulated using the standard Gillespie algorithm, along with the events in the population. We note that the rates $T_k^\pm$ need to be non-negative, i.e., we require $|m-k\Delta\sigma|<2v^2/\Delta\sigma$, for all $k$. In practice, this can be achieved by truncating the set of possible states $k\Delta \sigma$. More precisely, we disallow transitions out of the region $\{k: |m - k \Delta \sigma|\leq K\}$, with a given cutoff $K$. Provided that $K$ is sufficiently large truncations will only be required rarely. Once a cutoff $K$ is chosen we must require $\Delta \sigma \leq 2 v^2/K$ to guarantee non-negativity of the $T_k^\pm$. The variance of the Ornstein--Uhlenbeck process for $\sigma$ is given by $v^2$ in the long run [Eq.~(\ref{pst-sigma-eq})], so $K \propto v$ is a sensible choice. This results in maximum value for $\Delta \sigma$ which is also proportional to $v$. 

\subsubsection{Discrete-time simulation with explicit environmental dynamics (DEED)}\label{sec:discrete_time}

Approximate sample paths of the combined system of population and environment can also be generated in a discrete-time simulation. We refer to this as DEED (discrete-time simulation with explicit environmental dynamics). The time step $\Delta t$ needs to be sufficiently small to capture the details of the environmental process with characteristic time scale $\tau_c=\lambda^{-1}$. We therefore require $\Delta t \lesssim \lambda^{-1}$. One possible implementation is as follows:
\begin{enumerate}
\item[1.] Suppose we have arrived at time $t$, and the state of the population is $\bn(t)$ and that of the environment $\sigma(t)$. Obtain $\sigma(t+\Delta t)$ from Eq.~(\ref{ou-langevin}) using the Euler-Maruyama method \cite{maruyama1955continuous}.
\item[2.] Use $\sigma(t)$ and $\bn(t)$ to calculate the rates $p_r(t)=\Delta t \times R_{r,\sigma(t)}[\bn(t)]$ for $r=1,\dots,R$.
\item[3.] Provided $\Delta t$ is small enough, the $p_r(t)$ are all less than one. To lowest order in $\Delta t$ they are the probabilities that a reaction of type $r$ occurs in the next $\Delta t$. For each $r=1,\dots, R$ implement one reaction of this type with probability $p_r(t)$. With probability $1-p_r(t)$ no reaction of type $r$ occurs.  Executing all reactions that fire, one obtains $\bn(t+\Delta t)$. 
\item[4.] Increment time by $\Delta t$, and go to step 1.
\end{enumerate}
Step 3 disregards the possibility that a particular reaction fires multiple times during one time step. This is a valid approximation, provided that the $p_r(t)=\Delta t \times R_{r,\sigma(t)}$ are much smaller than one. As an alternative step 3 could be replaced by a conventional $\tau$-leaping step. The number of reactions of type $r$ that fire is then a Poissonian random variable with parameter $p_r(t)$.
\subsubsection{Thinning algorithm by Lewis}
A population subject to a dynamic external environment with continuous state space can also be simulated using the so-called thinning algorithm by Lewis \cite{lewis1979simulation}. This algorithm generates a statistically faithful ensemble of sample paths for Markovian systems with discrete states and transition rates with explicit external time dependence. 

In the context of our model the population is such a system. If the environmental dynamics is independent of the population then realisations $\sigma(t)$ for the environment can be generated in advance independently from the population.  For instance, sample solutions of the Ornstein-Uhlenbeck process in Eq.~(\ref{ou-langevin}) could be generated. Each such realisation $\sigma(t)$ then determines a realisation of time-dependent rates $R_r(\bn,t)\equiv R_{r,\sigma(t)}(\bn)$ for the population. The Lewis algorithm can then be used to produce sample paths for the population dynamics. 

In practice, numerical approximation schemes are required to generate realisations for the environment. For example, Eq.~(\ref{ou-langevin}) can be solved numerically using the Euler--Maruyama method, with time step $\Delta t$. As discussed above this time step needs to be sufficiently small ($\Delta t \lesssim \lambda^{-1}$) to resolve the short-time features of the environmental process. The Lewis algorithm then uses this as an input and generates sample paths for the population in continuous time.

\section{Application of the $\tau$FE to continuous-environmental models}\label{sec:contsim_examples}

In this section we test the $\tau$FE algorithm on a number of different examples of models with continuous environmental states. Simulation outcomes are compared against those from the algorithms described in Sec.~\ref{sec:contsim}.

\subsection{Toy model: Population dynamics with production and removal rates proportional to $\sigma^2$}\label{sec:sigsq}
 \begin{figure}[b]
  \centering
        \includegraphics[width=0.475\textwidth]{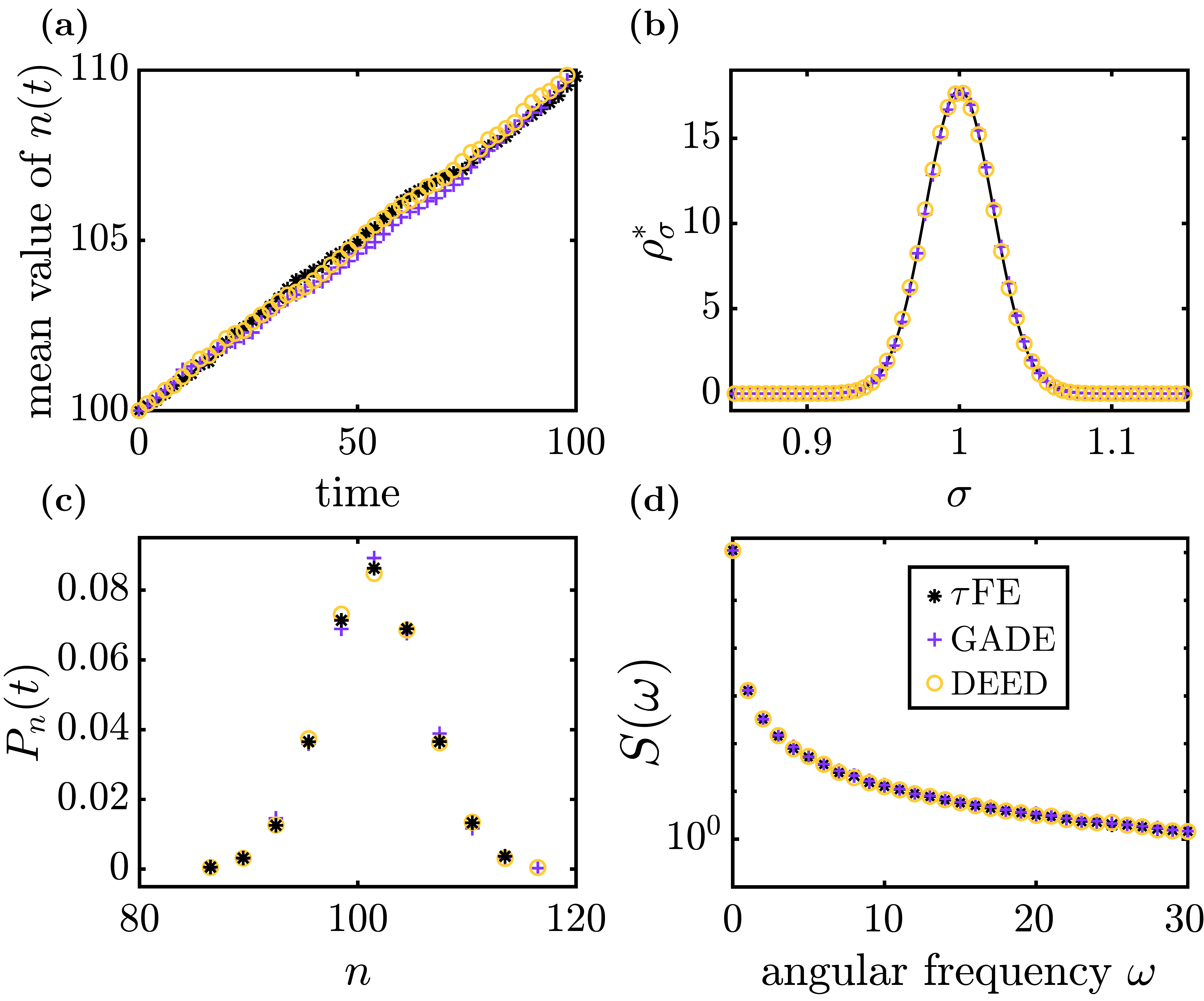}
    \caption{Simulation results for a production-removal process with rates $b = \beta \sigma^2$ and $d = \delta \sigma^2$ (Sec. \ref{sec:sigsq}), for the different algorithms described in Sec.~\ref{sec:continuous}. Parameters used: $\beta = 1.1$ and $\delta = 1.0$, $m=1$, $\lambda = 10^3$, and $v^2=5 \times 10^{-4}$. Panel (a): mean value of the number of individuals as function of time, obtained from $10^3$ runs. Panel (b): stationary distribution of the environmental state, $\rho_\sigma^*$, for the GADE method and the DEED approach (the $\tau$FE algorithm does not simulate the environment).  The solid line in panel (b) is the analytical solution from Eq.~(\ref{pst-sigma-eq}). Panel (c): distribution of the number of particles $n$ in the population at time $t=10$. Panel (d): spectral density [Eq.~(\ref{spectral-def})] obtained from $10^3$ runs. We use $\Delta \sigma = 10^{-3}$ for the GADE simulations, and $\Delta t =1/(100\lambda)=10^{-5}$ for DEED. For the $\tau$FE algorithm, we set $\Delta t = 10/\lambda=10^{-2}$.}
    \label{cont-env-birth-death}
\end{figure}
We first consider a production-removal process for a single species. The environmental state $\sigma(t)$ follows the Ornstein--Uhlenbeck process in Eq.~(\ref{ou-langevin}). The corresponding transition kernel $q_{\sigma\to\sigma'}(\tau)$ is given in Eq.~(\ref{eq:oukernel}), and the stationary distribution $\rho_\sigma^*$ in Eq.~(\ref{pst-sigma-eq}).  The production rate in the population is assumed to be $R_{b,\sigma} = \beta \sigma^2$, and the removal rate $R_{d,\sigma} = \delta \sigma^2$. These are not chosen with any particular natural system in mind, instead this example serves as an illustration (see also Appendix~\ref{app:f} for similar calculations for two related examples).
 
From (\ref{FSA-cont-avg}) we obtain
\BE
R_b^*  &=& \beta  (m^2 + v^2),\nonumber \\
R_d^*   &=& \delta  (m^2 + v^2).\label{eq:Rdbmean}
\EE
The second moments of the rates $\overline R_b(n)$ and $\overline R_d(n)$ can be calculated from Eq. (\ref{FSA-cont-second-mom}). We find 
\begin{align*}
& \avg{\overline R_b(n) \overline R_d(n)}-R_b^*(n)R_{d}^*(n) = \\
& \frac{\beta \delta  v^2 e^{-2 \lambda\Delta t}}{\lambda^2\Delta t^2} \Big[8 m^2  e^{\lambda\Delta t }+ \\ 
&e^{2 \lambda\Delta t} \left(8 m^2 (\lambda\Delta t-1 )+v^2 (2 \lambda\Delta t-1)\right)+ v^2\Big].\label{eq:abcov}\numberthis
\end{align*}
for the covariance. The expressions for the variances are similar, with suitable replacements $\beta\delta\to\beta^2$ and $\beta\delta\to\delta^2$ in the prefactor in Eq.~(\ref{eq:abcov}). This covariance matrix and the means in Eq.~(\ref{eq:Rdbmean}) are then used in the $\tau$FE algorithm.

\begin{table}[b]
\centering
\begin{tabular}{ m{4em}  m{6em}  m{6em}  m{6em} } 
\hline
\hline
\vspace{0.1em}
$\lambda^{-1}$ & GADE & DEED & $\tau$FE\\  
 \hline
$1 \times 10^{-2}$ &  28.47 & 3.84  & 0.79 $\times 10^{-2}$ \\ 
\hline
$5 \times 10^{-3}$ &  53.20 & 7.88  & 0.16 $\times 10^{-1}$ \\ 
\hline
$1 \times 10^{-3}$ & 288.30 & 40.91 &  0.08\\ 
\hline
$5 \times 10^{-4}$ & 576.69  & 82.78 & 0.15\\ 
\hline
$1 \times 10^{-4}$ &  3022.47  & 397.63 &  0.79\\ 
\hline
\hline
\end{tabular}
\caption{Mean computing time (in seconds) required for one simulation run of the model described in Sec.~\ref{sec:sigsq} until $t=10^3$. Data is from ten independent runs, parameters are as in Figure~\ref{cont-env-birth-death}, i.e., $\beta = 1.1, \delta = 1.0, m = 1,$ and $v^2=5 \times 10^{-4}$. For GADE we set $\Delta \sigma  = 10^{-3}$; for the DEED approach we set $\lambda\Delta t = 1/100$; for the $\tau$FE algorithm we set $\lambda\Delta t = 10$. }
\label{cont-env-bd-time-table}
\end{table}

Figure~\ref{cont-env-birth-death} shows simulation results from the $\tau$FE algorithm, as well as from the GADE and DEED schemes (Secs.~\ref{sec:gillespie_discrete} and \ref{sec:discrete_time} respectively). Panel (a) shows that all simulation methods result in linear growth (parameters are such that $\beta>\delta$, i.e., the growth rate is always larger than the death rate). Panel (b) confirms that GADE and DEED both generate the correct statistics for the stationary distribution of the environmental process [the solid line is the Gaussian distribution in Eq.~(\ref{pst-sigma-eq})].  In panel (c) we focus on a fixed time $t=10$, and show that all three simulation methods results in very similar distributions for the number of individuals in the population $n$ at that time.  Panel (d) finally shows a dynamic quantity, the Fourier spectrum $S(\omega)$ of the time series $n(t)$, or equivalently the Fourier transform of the correlation function of $n$. Again, all three simulation methods produce very similar results.

In Table \ref{cont-env-bd-time-table} we compare the average computing time required by the different algorithms to generate a trajectory up to time $t=10^3$. We show data for varying values of the typical time scale $\lambda^{-1}$ of the environmental process. GADE does not require any discretisation of time. For the DEED approach we use $\Delta t = 1/( 100 \lambda)$. For the $\tau$FE method we choose $\Delta t = 10/\lambda$. This is in-line with the requirements $\Delta t \lesssim \lambda^{-1}$ for DEED, and $\Delta t \gtrsim \lambda^{-1}$ for $\tau$FE. The choice of time steps will be discussed in further detail below.

The data in the table indicates that the simulation time scales approximately linearly with $\lambda$ for all three algorithms tested, provided $\lambda$ is sufficiently large. This is to be expected: The rates for the environmental events in the GADE simulations (Sec.~\ref{sec:gillespie_discrete}) scale as $\lambda$, and therefore dominate the events in the population for $\lambda \gg 1$. Each typical Gillespie step then advances time by an amount proportional to $\lambda^{-1}$, and ${\cal O}(\lambda)$ such steps are required to reach the designated end time. A similar argument applies to the DEED algorithm (Sec.~\ref{sec:discrete_time}) and for the $\tau$FE algorithm: For both of these we use time steps $\Delta t \propto \lambda^{-1}$, so again the number of iteration steps required scales as $\lambda$. 

The key message from Table~\ref{cont-env-bd-time-table} is that, for the choice of time steps made in the table, the computing time required by the $\tau$FE algorithm is substantially lower than that for the other two simulation methods. Given the linear dependence on $\lambda$, this increase in efficiency can be extrapolated to environments operating on time scales faster than the smallest time scale shown in the table (i.e., to the range $\lambda>10^{4}$). We note that, due to the smaller time step, DEED produces a finer resolution of sample paths in time than $\tau$FE. When we make our comparison we have average macroscopic quantities in mind (such as those in Fig.~\ref{cont-env-birth-death}), and not necessarily the generation of individual paths with the highest possible resolution in time. 
 
We now briefly discuss the choice of time steps for the $\tau$FE method and for DEED. In principle, we could have increased or decreased the step for either method. This would then reduce or increase the computing time required to reach the designated end point. It might also affect the accuracy of the outcome. Our choice of $\Delta t=10/\lambda$ for $\tau$FE is motivated by the good agreement with GADE in Fig.~\ref{cont-env-birth-death}, noting that GADE does not require any discretisation of time. Similarly, for the example discussed below in Sec.~\ref{sec:hill} good agreement with analytical predictions is found for this choice, see the regime of small $\lambda^{-1}$ in Fig.~\ref{assaf-PRL-tauhl}. Our conclusion is therefore that the $\tau$FE algorithm is able to produce results of the accuracy as in Fig.~\ref{cont-env-birth-death} with computing times as reported in Table~\ref{cont-env-bd-time-table}. 

The DEED algorithm requires $\Delta t\lesssim \lambda^{-1}$ to be able to resolve the environmental dynamics. Our choice $\Delta t = 1/(100\lambda)$ in Table~\ref{cont-env-bd-time-table} is well below this requirement, and the algorithm can in principle be speed up by choosing a larger time step. If we were to exhaust the limit and used $\Delta t = \lambda^{-1}$ for DEED then this would reduce the computing time by about a factor of one hundred in Table~\ref{cont-env-bd-time-table}. For for $\lambda^{-1}=10^{-3}$ this would mean a reduction from approximately $40$ seconds to $0.4$ seconds per sample path. Using this larger time step also results in noticeable deviations in measurements of the quantities in Fig.~\ref{cont-env-birth-death} from continuous-time GADE simulations. But even if we accept this and use the hundred fold larger time step for DEED the $\tau$FE algorithm would remain approximately five times faster, requiring $0.08$ seconds per sample path at $\lambda^{-1}=10^{-3}$, see Table~\ref{cont-env-bd-time-table}. 

We have also conducted tests with Lewis' thinning algorithm. To do this we have first generated sample paths of the Ornstein--Uhlenbeck process for the environment [Eq.~(\ref{ou-langevin})] using an Euler--Maruyama scheme. This is then fed into the Lewis' algorithm for systems with time dependent rates. Given that the typical time scale of the environment is $\lambda^{-1}$,  the largest sensible time step for the Euler--Maruyama scheme is $\Delta t = \lambda^{-1}$, similar to DEED. This choice minimises the computing time for the Lewis' approach. We therefore use this time step to compare the  efficiency of the Lewis' approach with that of $\tau$FE. We find that the thinning algorithm is considerably slower than the $\tau$FE approach. For $\lambda^{-1} = 10^{-3}$, for example, we obtain a simulation time of approximately $13$ seconds per run up to $t=10^3$ compared to $0.08$ seconds for $\tau$FE (see Table~\ref{cont-env-bd-time-table}).

\subsection{Genetic switch with Hill-like regulatory function}\label{sec:hill}
As a final example we consider a model of protein production subject to a continuous environment discussed in \cite{assaf2013extrinsic}. The model entails positive feedback, in that the presence of protein has the potential to increase production of protein. There is one single species in the model (protein), we write the number of protein molecules as $n$. We also define $x=n/\Omega$, where $\Omega$ is again a model parameter setting the typical size of the system. The production rate of protein is given by
\begin{equation}\label{eq:prod}
f(x, \sigma) = \alpha_0 + (1 - \alpha_0 + \sigma) \Theta(x - x_0),
\end{equation}
where $0<\alpha_0<1$ and $x_0>0$ are constants, and where $\Theta(x)$ is the Heaviside function. Protein molecules also decay with unit rate. 
In the absence of environmental influence ($\sigma\equiv 0$), the production rate is thus unity when $x>x_0$, and $\alpha_0<1$ when $x<x_0$. For $\sigma\equiv 0$ the mean re-scaled number of protein follows the rate equation
\begin{equation}\label{eq:xdot}
\dot{\bar{x}} = f(\bar{x}) - \bar{x},
\end{equation}
where time is measured in units of generations. We choose $\alpha_0<x_0<1$. Eq.~(\ref{eq:xdot}) has three fixed points $x_1^* < x_2^* < x_3^*$, where $x_1^* = \alpha_0$ and $x_3^* = 1$ are attractors, and $x_2^* = x_0$ is a repeller.  Similar to \cite{assaf2013extrinsic}, we refer to $x_1^*$ and $x_3^*$ as the `low' and `high' states, respectively.

 The environmental process $\sigma(t)$ modulates the production rate when $x>x_0$. As in \cite{assaf2013extrinsic} we asssume that $\sigma$ follows an Ornstein-Uhlenbeck process of the form given in Eq.~(\ref{ou-langevin}). The noisy system has the potential to switch between the `high' and `low' states. To test the performance of the $\tau$FE algorithm, we focus on the mean switching time (MST)  to transit from the high state to the low state. This time is studied and calculated in \cite{assaf2013extrinsic}, we denote it by $\avg{\tau_{\text{high} \rightarrow \text{low}}}$. In simulations we start the system in the high state, and measure the first time the system reaches the low state.

Only the production of protein is affected by the state $\sigma$ of the environment, we write $R_{{\rm prod},\sigma}(\bn) = f(x,\sigma)$, with $f$ as in Eq.~(\ref{eq:prod}). Inserting this in Eqs.~(\ref{FSA-cont-avg}) and (\ref{FSA-cont-second-mom}), and after straightforward calculations, we obtain 
\begin{equation}
R_{\rm prod}^*(n) =  \alpha_0 + (1 - \alpha_0) \Theta(n/\Omega - x_0),
\end{equation}
and the second moment 
\begin{align*}
& \avg{(\overline R_{\rm prod}(n))^2} -[R_{\rm prod}^*(n)]^2  = \\
 & \dfrac{2 v^2 }{\lambda^2\Delta t^2} \left[ \lambda\Delta t +  (e^{-\lambda\Delta t} - 1) \right] \Theta(n/\Omega - x_0).\numberthis
\end{align*}

 \begin{figure}[t!!]
  \centering
        \includegraphics[width=0.475\textwidth]{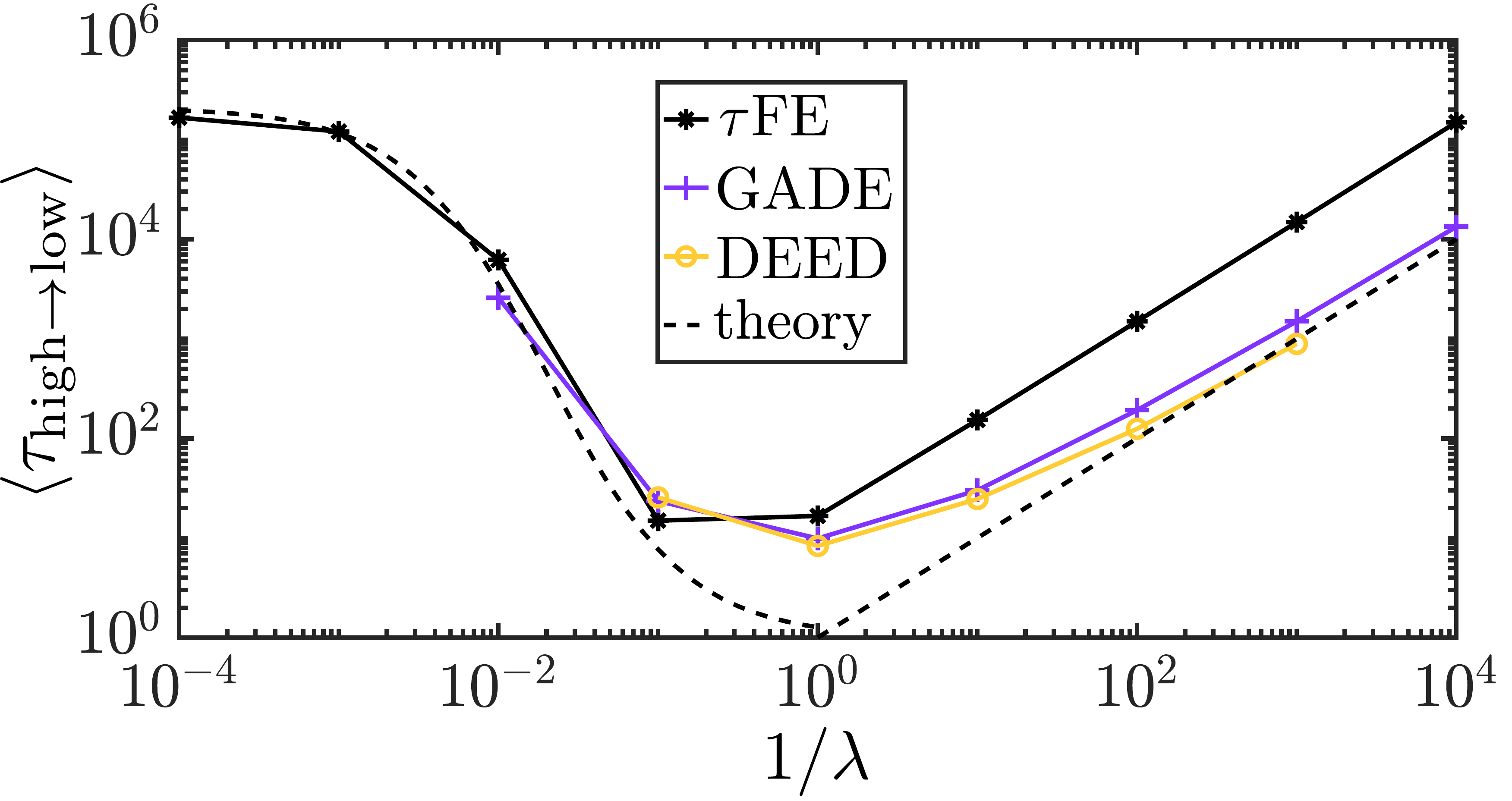}
    \caption{Mean switching time (MST) from the high to the low state in the model described in Sec.~\ref{sec:hill}. For the $\tau$FE algorithm we used $\Delta t = 10 /\lambda$, for GADE $\Delta \sigma = 0.01$, and for DEED $\Delta t = 10^{-4}$. Theory curves are from Eqs.~(8) and (10) in \cite{assaf2013extrinsic}. Model parameters are as in the top right panel of Fig.~2  in~\cite{assaf2013extrinsic} ($\Omega=5000, v = 0.1, \alpha_0 = 0.01, x_0 = 0.93$).}
    \label{assaf-PRL-tauhl}
\end{figure}

In Fig.~\ref{assaf-PRL-tauhl} we show the MST measured in simulations using the different approaches described in in Sec.~\ref{sec:continuous}. Assaf et al. \cite{assaf2013extrinsic} report non-monotonous behaviour of the MST as a function of $\tau_c=\lambda^{-1}$. As seen in Fig.~\ref{assaf-PRL-tauhl} the $\tau$FE algorithm reproduces this behaviour. For fast environmental dynamics (low $\lambda^{-1}$) the MST obtained from the $\tau$FE algorithm is in good agreement with measurements obtained from the other simulation methods, and with the analytical approximations from \cite{assaf2013extrinsic}. The agreement extends over several decades of values of $\tau_c=\lambda^{-1}$.

At the same time we observe that the $\tau$FE algorithm requires significantly less computing time than the GADE or DEED approaches.  For $\lambda^{-1}=10^{-1}$ for example, we measured an average computing time of $2\times 10^{-3}$ seconds to generate one run of the system up to time $10^3$ with the $\tau$FE algorithm ($\Delta t=10/\lambda$). GADE required $0.674$ seconds, and DEED $4.2$ seconds (for a time step $\Delta t = 10^{-4}$). 

We note that we have implemented DEED as described in Sec.~\ref{sec:discrete_time}. In particular at most one reaction of each type can fire in each time step (step 3 of the algorithm). This requires a sufficiently small time step $\Delta t$ to ensure $p_r(t)<1$ for all $r$.  This is achieved by our choice $\Delta t = 10^{-4}$. Alternatively step 3 of the DEED algorithm could be replaced by a  (conventional) $\tau$-leaping step. Larger choices of the time step $\Delta t$ are then possible, up to the limit of $\Delta t \approx \lambda^{-1}$ to ensure that the environmental dynamics are captured appropriately. Focusing on $\lambda^{-1}=10^{-1}$ we expect that increasing the time step by a factor of a thousand (from $10^{-4}$ to $10^{-1}$) would reduce the simulation time by at most a factor of a thousand for a $\tau$-leaping version of DEED. This would result in a computing time of approximately $4\times 10^{-3}$ for one simulation run up to $t=10^3$ instead of the $4.2$ seconds reported for DEED in the previous paragraph. This is comparable with the CPU time required by the $\tau$FE algorithm ($2\times 10^{-3}$ seconds), but would resolve environmental fluctuations with lower accuracy. For example one observes systematic deviations for the stationary distribution of the environment in Fig.~\ref{cont-env-birth-death}(b).

\section{Discussion and conclusions} \label{sec:conclusions}
In summary, we have presented $\tau$FE, a variant of the $\tau$-leaping stochastic simulation algorithm for systems subject to fast environmental dynamics. Just like conventional $\tau$-leaping the algorithm operates in discrete time. The rates of the reactions in the system proper are treated as constant during each time step, and the numbers of different reactions firing in one step have Poissonian statistics. 

The key difference compared to conventional $\tau$-leaping is the external environment. In the full continuous-time model reaction rates which depend on the environmental state fluctuate in time even when the state of the population does not change. An adiabatic approximation would consist of assuming an infinitely fast environment and of replacing the reaction rates by their means with respect to the stationary distribution of the environmental process. This is justified if the relaxation time scale of the environmental process is infinitely shorter than the time step of the simulation. 

The $\tau$FE algorithm goes beyond this approximation, and is based on time averages of reaction rates over the {\em finite} time step. For finite speeds of the environment these average rates are random variables. If the environmental dynamics is fast we can make a Gaussian approximation. The rates feeding into the $\tau$-leaping step are clipped Gaussian random numbers designed to retain the first and second moments of the actual environmental dynamics. It is important to note that this not the same as drawing an environmental state $\sigma$ from the stationary distribution $\rho^*_\sigma$, and then using the rates $R_{r,\sigma}(\bn)$ for the next $\tau$-leaping step. Instead, the covariance matrix of the rates $\overline R_r(\bn)$ in Eq.~(\ref{eq:rbar}) is calculated as described in Eqs.~(\ref{avg-prod-Rs}) for discrete environments, and in Eq.~(\ref{FSA-cont-second-mom}) for  continuous environmental states.

The choice of time step for the $\tau$FE algorithm requires careful consideration. On the one hand the time step must be long enough to justify the averaging procedure over the environmental dynamics and the Gaussian assumption for the reaction rates in the $\tau$-leaping step. Broadly speaking $\lambda\Delta t$ must be sufficiently large ($\lambda\Delta t \gg 1$). At the same time the so-called leap condition for the $\tau$-leaping part of the algorithm must be fulfilled \citep{Gillespie_2001}. This means that the state of the system must not change significantly in each iteration step, as a constant state $\bn$ of the population is an assumption made in setting up the $\tau$-leaping. Mathematically, this means that the change of the number of particles in the system in a time step must be much smaller than the typical number of particles in the system. Assuming that the stoichiometric coefficients do not scale with the system size $\Omega$ this means that $ \Delta t \times R_{r,\sigma}(\bn)$ must be much smaller than $\Omega$. Noting that $R_{r,\sigma}(\bn)$ is of order $\Omega$ in many applications we thus require that $\Delta t$ is much smaller than one. For $\lambda\gg1$ and $\Delta t$ proportional to $\lambda^{-1}$ this condition is often relatively easy to meet in practice.

We have tested the $\tau$FE algorithm on a number of systems with discrete and continuous environments. This includes examples of systems which can be addressed analytically and models motivated by applications in biology. Our tests focus on  stationary distributions, but also dynamic features such as Fourier spectra of fluctuations or first-passage time distributions. In all cases we have tested the $\tau$FE method produces good agreement with results from conventional simulation methods in the regime of fast environmental dynamics. This is the regime for which $\tau$FE is designed. Naturally, quantitative deviations are found when the time scales of the environmental dynamics and system proper are insufficiently separated.

We stress that $\tau$FE goes beyond simulations in the adiabatic limit, and is able to capture the dependence of macroscopic observables on the time scale separation, provided this dependence is sufficiently strong [see e.g. Figs.~\ref{waiting-times-bim-gen-swt}(d) and \ref{assaf-PRL-tauhl})]. At the same time our analysis also reveals limitations of the algorithm. If the dependence of observables on the time scale separation is weak such as in Fig.~\ref{waiting-times-bim-gen-swt}(c), then $\tau$FE may not be able to fully resolve these dependencies. When the environment is fast the quantitative agreement with simulations of the full system is however still within approximately $10\%$ in the example in Fig.~\ref{waiting-times-bim-gen-swt}(c).

The computing time required for the $\tau$FE algorithm to generate sample paths up to a designated end time is proportional to the inverse time step. The time step on the other hand is typically a multiple of the characteristic time scale $\lambda^{-1}$ of the environmental dynamics. This means that the computational effort scales approximately linearly in the time scale separation $\lambda$. In all cases we have tested we found that $\tau$FE is considerably more efficient for the measurement of macroscopic quantities than alternative simulation algorithms. 

In summary, we think the $\tau$FE algorithm has passed the initial selection of tests presented in this paper. It provides an promising approach to probing the regime of fast environmental dynamics, and captures effects induced by extrinsic noise beyond the adiabatic limit. The algorithm is particularly valuable for systems in which the regime of intermediate time scale separation can be accessed with conventional simulation methods. The accuracy of the $\tau$FE algorithm can then be assessed in this regime (an example can be found in Fig.~\ref{assaf-PRL-tauhl}). If the comparison is favourable, then it is justified to use $\tau$FE in the regime of increasing time scale separation.

\section*{Acknowledgements}

We would like to thank Yen Ting Lin (Los Alamos) for useful discussions and feedback on earlier versions of the manuscript. EBC acknowledges a President's Doctoral Scholarship (The University of Manchester). TG acknowledges funding from the Spanish Ministry of Science, Innovation and Universities, the Agency AEI and FEDER (EU) under the grant PACSS (RTI2018-093732-B-C22), and the Maria de Maeztu program for Units of Excellence in R\&D (MDM-2017-0711).

\bibliography{tau-leaping-ref}

\onecolumngrid

\begin{appendix}
\section{Second moments of rates}\label{app:second_moment}
In this Appendix we calculate the second moments of the quantities $\overline R_r(\bn)$ ($r=1,\dots,R$) defined in Eq.~(\ref{eq:rbar}). Without loss of generality we assume that the time interval in question starts at $t=0$, the end point is then $\Delta t$.
Assuming the space of environmental states is discrete, we have
\BE
\avg{\overline R_r(\bn) \overline R_s(\bn)}&=&
\frac{1}{\Delta t^2}\int_0^{\Delta t} \mathrm{d} t_1 \int_0^{\Delta t}  \mathrm{d} t_2 \avg{R_{r,\sigma(t_1)}(\bn)R_{s,\sigma(t_2)}(\bn)} \nonumber \\
&=&\frac{1}{\Delta t^2}\int_0^{\Delta t} \mathrm{d}t_1 \int_{t_1}^{\Delta t}\mathrm{d}t_2 \avg{R_{r,\sigma(t_1)}(\bn)R_{s,\sigma(t_2)}(\bn)}+\frac{1}{\Delta t^2}\int_0^{\Delta t} \mathrm{d}t_2 \int_{t_2}^{\Delta t} \mathrm{d}t_1 \avg{R_{r,\sigma(t_1)}(\bn)R_{s,\sigma(t_2)}(\bn)}
\nonumber \\
&=&\frac{1}{\Delta t^2}\sum_{\sigma\sigma'}  \int_0^{\Delta t} \mathrm{d}t_1 \int_{t_1}^{\Delta t} \mathrm{d}t_2 ~\rho^*_\sigma q_{\sigma\to\sigma'}(t_2-t_1)R_{r,\sigma}(\bn)R_{s,\sigma'}(\bn)\nonumber \\
&&+\frac{1}{\Delta t^2}\sum_{\sigma\sigma'}  \int_0^{\Delta t} \mathrm{d}t_2 \int_{t_2}^{\Delta t} \mathrm{d}t_1 ~\rho^*_{\sigma'} q_{\sigma'\to\sigma}(t_1-t_2)R_{r,\sigma}(\bn)R_{s,\sigma'}(\bn)\nonumber \\
&=&\frac{1}{\Delta t^2}\sum_{\sigma\sigma'}  \int_0^{\Delta t} \mathrm{d}t_1 \int_{t_1}^{\Delta t} \mathrm{d}t_2 ~\rho^*_\sigma q_{\sigma\to\sigma'}(t_2-t_1)R_{r,\sigma}(\bn)R_{s,\sigma'}(\bn)\nonumber \\
&&+\frac{1}{\Delta t^2}\sum_{\sigma\sigma'}  \int_0^{\Delta t} \mathrm{d}t_1 \int_{t_1}^{\Delta t} \mathrm{d}t_2~ \rho^*_{\sigma} q_{\sigma\to\sigma'}(t_2-t_1)R_{r,\sigma'}(\bn)R_{s,\sigma}(\bn).\label{eq:help1}
\EE
In the first step we have applied the definition of the over-bar average [Eq.~(\ref{eq:rbar})]. In the third step we have carried out the average over realisations of the environmental process. In the last step we have renamed $t_1\leftrightarrow t_2$ and $\sigma\leftrightarrow\sigma'$ in the second term. Therefore
\BE
\avg{\overline R_r(\bn) \overline R_s(\bn)}
&=&\frac{1}{\Delta t^2}\sum_{\sigma\sigma'}  \int_0^{\Delta t} \mathrm{d}t_1 \int_{t_1}^{\Delta t} \mathrm{d}t_2~ \rho^*_\sigma q_{\sigma\to\sigma'}(t_2-t_1) \big[R_{r,\sigma}(\bn)R_{s,\sigma'}(\bn)+R_{r,\sigma'}(\bn)R_{s,\sigma}(\bn)\big].
\label{avg-prod-Rs-app}
\EE
Up to a shift of the start point of the time step, this is identical to Eq.~(\ref{avg-prod-Rs}).

\medskip

As explained in Section~\ref{sec:implem-algo}, the sums over $\sigma$ become integrals when the environment takes continuous states. We then find Eq.~(\ref{FSA-cont-second-mom}). 

When the environmental space is discrete, we can use Eq.~(\ref{eq:q}) and find
\begin{align*}
\avg{\overline R_r(\bn) \overline R_s(\bn)}
&=\frac{1}{\Delta t^2}\sum_{\sigma\sigma'}  \int_0^{\Delta t} \mathrm{d}t_1 \int_{t_1}^{\Delta t} \mathrm{d}t_2 ~\rho^*_\sigma \rho^*_{
\sigma'} \big[R_{r,\sigma}(\bn)R_{s,\sigma'}(\bn)+R_{r,\sigma'}(\bn)R_{s,\sigma}(\bn)\big] \\
&+\frac{1}{\Delta t^2}\sum_{\sigma\sigma'}\sum_{\ell=2}^M  \int_0^{\Delta t} \mathrm{d}t_1 \int_{t_1}^{\Delta t} \mathrm{d}t_2 ~\rho^*_\sigma c_{\ell,\sigma} v_{\ell,\sigma'}e^{-\lambda\mu_\ell (t_2- t_1)}  \big[R_{r,\sigma}(\bn)R_{s,\sigma'}(\bn)+R_{r,\sigma'}(\bn)R_{s,\sigma}(\bn)\big]\\
&= R_{r,\rm avg}(\bn)R_{s,\rm avg}(\bn)\\
& +\frac{1}{\Delta t^2}\sum_{\sigma\sigma'}\sum_{\ell=2}^M\rho^*_\sigma c_{\ell,\sigma} v_{\ell,\sigma'} \big[R_{r,\sigma}(\bn)R_{s,\sigma'}(\bn)+R_{r,\sigma'}(\bn)R_{s,\sigma}(\bn)\big] \int_0^{\Delta t} \mathrm{d}t_1 \int_{t_1}^{\Delta t} \mathrm{d}t_2~ e^{\lambda\mu_\ell (t_2-t_1)}. \numberthis
\end{align*}

\section{Further details for systems with two species and two environmental states}
\label{app:two-sp-two-env}

The case of two species and two environmental states ($S=2, M=2$) was studied in \cite{hufton2019classical}, and a simple version of the $\tau$FE algorithm was presented for this restricted case. We assume $\sigma$ switches from state $0$ to state $1$ with rate $\lambda k_1$, and from $1$ to $0$ with rate $\lambda k_0$. The environmental transition matrix then becomes
\be
\bA=\left(\begin{array}{cc} -k_1 & k_0 \\ k_1 & -k_0\end{array}\right),
\ee
whose eigenvalues are $\mu_1 = 0$ and $\mu_2=-(k_0+k_1)$. The respective eigenvectors take the form
\be
\bv_1 = \brho^*=\frac{1}{k_0+k_1}\left(\begin{array}{c} k_0 \\ k_1 \end{array}\right) \quad \text{and} \quad \bv_2=\left(\begin{array}{c} 1 \\ -1 \end{array}\right),
\ee
where $\brho^*$ has been normalised to represent the stationary distribution for $\sigma$. The coefficients $c_{2,0}$ and $c_{2,1}$ are obtained from Eq.~(\ref{eq:coeff}), for the initial conditions $\brho(0) = (1,0)$ and $\brho(0) = (0,1)$. We find 
\be
c_{2,0}=\frac{k_1}{k_0+k_1} \quad \text{and} \quad c_{2,1}=\frac{-k_0}{k_0+k_1}.
\ee
Putting all together in Eq.~(\ref{eq:rr}), and after straightforward calculations we arrive at
\BE
\Xi_{rs}\equiv \avg{\overline R_r(\bn) \overline R_s(\bn)}-R_{r}^*(\bn)R_{s}^*(\bn) =\frac{\theta^2}{\lambda\Delta t} \left[ R_{r,1}(\bn)-R_{r,0}(\bn)\right]\left[ R_{s,1}(\bn)-R_{s,0}(\bn)\right],
\EE
where $\theta^2 = 2 k_0 k_1/(k_0 + k_1)^3$. The indices $r$ and $s$ stand for reactions affected by the environment. As explained in Section~\ref{sec:algo-description}, to simulate the $\tau$FE algorithm we need to draw correlated Gaussian random numbers $\overline R_r$ with means
\begin{equation}
R^*_r(\bn) = \dfrac{k_0 R_{r,0} + k_1 R_{r,0}}{k_0 + k_1},
\end{equation}
for $r=1,2$, and covariance matrix
\be
\mathbf{\Sigma}=\left(\begin{array}{cc} \Xi_{11} & \Xi_{12} \\ \Xi_{21} & \Xi_{22} \end{array}\right).
\label{cov-matrix-def}
\ee
One way to do this is by drawing independent Gaussian random numbers $z_1$ and $z_2$ with mean zero and unit variance, and then to set
\be
\left(\begin{array}{c} \overline R_1(\bn) \\ \overline R_2(\bn) \end{array}\right) = \left(\begin{array}{c} R_1^*(\bn) \\  R_2^* (\bn)\end{array}\right) + \mathbf{C} \left(\begin{array}{c} z_1 \\ z_2 \end{array}\right),
\label{corr-gaussian}
\ee
with a matrix $\mathbf{C}$ that fulfils $\mathbf{C} \mathbf{C}^T = \mathbf{\Sigma}$, where $T$ denotes the transpose. This matrix is not unique. We use
\be
\mathbf{C} = \dfrac{\mathbf{\Sigma}}{\sqrt{\theta^2/(\lambda \Delta t) \left \{\left[ R_{1,1}(\bn)-R_{1,0}(\bn)\right]^2 + \left[ R_{2,1}(\bn)-R_{2,0}(\bn)\right]^2 \right \}}}.
\ee

\section{Birth-death process with two species and three environmental states}
\label{app:three-env-two-sp}

In the example in Sec.~\ref{sec:ex2} we have the following transition matrix for the environmental process
\begin{equation}
\bA=\left(\begin{array}{ccc} -k_1 & 0 & k_0 \\ k_1 & -k_2 & 0 \\ 0 & k_2 & -k_0 \end{array}\right).
\end{equation}
The eigenvalues of this matrix are
\be
\mu_1 = 0, \quad \mu_2 = -\frac{1}{2} \left( k_0 + k_1 + k_2 + \Gamma \right), \quad \text{and}, \quad \mu_3 = -\frac{1}{2} \left( k_0 + k_1 + k_2 - \Gamma \right),
\ee 
with $\Gamma = \sqrt{k_0^2 + k_1^2 + k_2^2 -2(k_0 k_1 + k_1 k_2 + k_2 k_0)}$. The associated eigenvectors take the form
\be
\bv_1 = \brho^*=\frac{1}{k_0 k_1 + k_1 k_2 + k_2 k_0}\left(\begin{array}{c} k_2 k_0 \\ k_0 k_1 \\ k_1 k_2 \end{array}\right),
\ee
and
\be
\bv_2=\left(\begin{array}{c} (-k_0+k_1-k_2 + \Gamma)/(2 k_2) \\ (k_0-k_1-k_2 - \Gamma)/(2 k_2)\\ 1\end{array}\right), \quad \quad \bv_3=\left(\begin{array}{c}(-k_0+k_1-k_2 - \Gamma)/(2 k_2) \\ (k_0-k_1-k_2 + \Gamma)/(2 k_2) \\ 1\end{array}\right).
\ee
Using Eq.~(\ref{eq:coeff}) and three sets of initial conditions (each concentrated on one environmental state) we find
\be
c_{2,0} = \dfrac{k_1 k_2 \left( k_0 + k_1 + k_2 - \Gamma \right)}{2 \Gamma(k_0 k_1 + k_1 k_2 + k_2 k_0)}, \quad c_{3,0} = -\dfrac{k_1 k_2 \left( k_0 + k_1 + k_2 + \Gamma \right)}{2 \Gamma(k_0 k_1 + k_1 k_2 + k_2 k_0)},
\ee
as well as
\be
c_{2,1} = -\dfrac{k_2 \left( k_0( k_1 + 2 k_2) + k_1 \left( - k_1 + k_2 + \Gamma \right) \right)}{2 \Gamma(k_0 k_1 + k_1 k_2 + k_2 k_0)}, \quad c_{3,1} = \dfrac{k_2 \left( k_0( k_1 + 2 k_2) + k_1 \left( -k_1 + k_2 - \Gamma \right) \right)}{2 \Gamma(k_0 k_1 + k_1 k_2 + k_2 k_0)},
\ee
and finally
\be
c_{2,2} = \dfrac{k_0 \left( -k_1^2 - k_2^2 + k_0 (k_1 + k_2) + k_1 \Gamma + k_2 \Gamma \right)}{2 \Gamma(k_0 k_1 + k_1 k_2 + k_2 k_0)}, \quad c_{3,2} = \dfrac{k_0 \left( k_1^2 + k_2^2 - k_0 (k_1 + k_2) + k_1 \Gamma + k_2 \Gamma \right)}{2 \Gamma(k_0 k_1 + k_1 k_2 + k_2 k_0)}.
\ee
Putting all together in Eqs.~(\ref{eq:rstar}) and (\ref{eq:rr}) and after further tedious but straightforward calculations, we arrive at the expressions in Eqs.~(\ref{alpha-beta-star}) and (\ref{cov-three-env}).

In order to draw the correlated Gaussian random numbers $\bar{\alpha}$ and $\bar{\beta}$ required for the $\tau$-leaping step, we proceed as in Appendix~\ref{app:two-sp-two-env}. We construct the covariance matrix $\mathbf{\Sigma}$ [Eq.~(\ref{cov-matrix-def})] and then find a matrix $\mathbf{C}$ such that $\mathbf{C} \mathbf{C}^T = \mathbf{\Sigma}$. We then draw independent Gaussian random numbers $z_1$ and $z_2$ with mean zero and unit variance, and use an expresion analogous to that in Eq.~(\ref{corr-gaussian}) to obtain $\bar{\alpha}$ and $\bar{\beta}$. The matrix $\mathbf{C}$ we use is
\be
\mathbf{C} = A \left(\begin{array}{cc} \dfrac{\sigma_{\alpha \alpha} + B}{\sigma_{\alpha \beta}} & 1 \\ 1 & \dfrac{\sigma_{\beta \beta} + B}{\sigma_{\alpha \beta}} \end{array}\right),
\ee
with $\sigma_{\alpha\alpha}$ and $\sigma_{\alpha\beta}$ as given in Eq.~(\ref{cov-three-env}), and
\be
A = \dfrac{\sigma_{\alpha \beta}}{\sqrt{\sigma_{\alpha \alpha} + \sigma_{\beta \beta} + B}},
\ee
and
\BE
B = \dfrac{\sqrt{3} k_0 k_1 k_2}{\lambda \Delta t (k_0 k_1 + k_1 k_2 + k_2 k_0)^2} \times \abs{\alpha_0(\beta_2 - \beta_1) + \alpha_1(\beta_0 - \beta_2) + \alpha_2 (\beta_1 - \beta_0)}.
\EE

\section{Bimodal genetic switch}
\label{app:bim-gen-swt}

For the model in Sec.~\ref{sec:ex3} the rates of the environmental transitions depend on the number of proteins $n_P$ in the population. We assume that $n_P$ remains constant during each $\tau$-leaping step. The environmental transition matrix then becomes
\begin{equation}
\bA=\left(\begin{array}{ccc} -\tilde{k}_+ & k_- & 0 \\ \tilde{k}_+ & -\tilde{k}_+ - k_- & k_- \\ 0 & \tilde{k}_+ & -k_- \end{array}\right),
\end{equation}
with $\tilde{k}_+ = k_+ n_P/\Omega$. The eigenvalues of this matrix are
\BE
\mu_1 = 0,\quad
\mu_2 = -k_- - \tilde{k}_+ - \sqrt{k_- \tilde{k}_+}, \quad
 \mu_3 = -k_- - \tilde{k}_+ + \sqrt{k_- \tilde{k}_+},
\EE
while the associated eigenvectors take the form
\be
\bv_1 = \brho^*=\frac{1}{k_-^2 + k_- \tilde{k}_+ +  \tilde{k}_+^2}\left(\begin{array}{c} k_-^2\nonumber \\ k_- \tilde{k}_+ \\ \tilde{k}_+^2 \end{array}\right),\ee
\BE
\bv_2=\left(\begin{array}{c} \sqrt{k_-/\tilde{k}_+} \\  \left(-\sqrt{k_-} - \sqrt{\tilde{k}_+} \right)/\sqrt{\tilde{k}_+}\\ 1\end{array}\right), \quad \text{and,} \quad \bv_3=\left(\begin{array}{c} -\sqrt{k_-/\tilde{k}_+} \\  \left(\sqrt{k_-} - \sqrt{\tilde{k}_+} \right)/\sqrt{\tilde{k}_+}\\ 1\end{array}\right).
\EE
Applying Eq.~(\ref{eq:coeff}) for different initial conditions as above, we obtain
\be
c_{2,0} = \dfrac{\tilde{k}_+^{3/2}}{2 \sqrt{k_-} \left(k_- + \sqrt{k_- \tilde{k}_+} + \tilde{k}_+ \right)},
\quad
c_{3,0} = -\dfrac{\tilde{k}_+^{3/2}}{2 \sqrt{k_-} \left(k_- - \sqrt{k_- \tilde{k}_+} + \tilde{k}_+ \right)},
\ee
as well as
\be
c_{2,1} = -\dfrac{\tilde{k}_+ + \sqrt{k_- \tilde{k}_+ }}{2 \left(k_- + \sqrt{k_- \tilde{k}_+} + \tilde{k}_+ \right)}, 
\quad
 c_{3,1} = -\dfrac{\tilde{k}_+ - \sqrt{k_- \tilde{k}_+ }}{2 \left(k_- - \sqrt{k_- \tilde{k}_+} + \tilde{k}_+ \right)},
\ee
and finally
\be
c_{2,2} = -\dfrac{k_-}{2 \left(k_- + \sqrt{k_- \tilde{k}_+} + \tilde{k}_+ \right)},
\quad 
c_{3,2} = -\dfrac{k_-}{2 \left(k_- - \sqrt{k_- \tilde{k}_+} + \tilde{k}_+ \right)}.
\ee
Putting this together in Eqs.~(\ref{eq:rstar}) and (\ref{eq:rr}) and after straightforward calculations, we arrive at the expressions in Eqs.~(\ref{b-star-bim-gen-swt}) and (\ref{cov-bim-gen-swt}).

Since only one reaction is affected by the environmental state, it is only necessary to drawn one Gaussian random number with mean $b^*$ and variance $\sigma_{bb}$ in each step of the $\tau$FE algorithm, with $b^*$ and $\sigma_{bb}$ given in Eqs.~(\ref{b-star-bim-gen-swt}) and (\ref{cov-bim-gen-swt}) respectively.

\section{Gillespie algorithm with discretised environmental dynamics (GADE)} 
\label{app:gillespie-cont}
In this Appendix we briefly describe the constructions of the rates given in Eq.~(\ref{discrete-sigma-rates}). They define a continuous-time dynamics on a discrete state space approximating the Ornstein--Uhlenbeck process in Eq.~(\ref{ou-langevin}).

\medskip

\noindent{\em Matching the first moments of movements.} We first look at the mean drift of $\sigma$, i.e., the mean change of $\sigma$ per unit time. Suppose the environment is in a given state $\sigma$. The mean drift in the Ornstein--Uhlenbeck process [Eq.~(\ref{ou-langevin})] is then $\lambda(m-\sigma)$. 

\medskip
Suppose now the above discrete-$\sigma$ process is in state $\sigma=k\Delta\sigma$. Then $\sigma$ increases to $\sigma+\Delta\sigma$ with rate $T_k^+$ and decreases to $\sigma-\Delta\sigma$ with rate $T_k^-$. The expected change (per unit time) is therefore $\Delta \sigma \times (T_k^+-T_k^-)$. 

We conclude that we need to impose
\be\label{eq:aux1}
\Delta\sigma \times (T_k^+-T_k^-)=\lambda(m-k\Delta\sigma).
\ee

\medskip

\noindent{\em Matching the variance of movements.} Next we look at the variance of movements of $\sigma$. For the Ornstein--Uhlenbeck process in Eq.~(\ref{ou-langevin}) the second moment of movements (per unit time) is given by $2 \lambda v^2$. In the discrete-$\sigma$ process, the second moment of movements is $(\Delta \sigma)^2\times (T_k^+ + T_k^-)$. To match the Ornstein--Uhlenbeck process, we then need to impose
\be\label{eq:aux2}
(\Delta \sigma)^2\times (T_k^+ + T_k^-)=2 \lambda v^2.
\ee

\noindent{\em Overall solution.} Simultaneously solving Eqs.~(\ref{eq:aux1}) and (\ref{eq:aux2}) for $T_k^+$ and $T_k^-$ we arrive at Eq.~(\ref{discrete-sigma-rates}).

\section{Additional examples of production-removal processes in continuous environments}\label{app:f}

In this Appendix we include results for the variances and covariances $\avg{\overline R_r(n) \overline R_s(n)}-R_{r}^*(\bn)R_{s}^*(n)$ for two further exemplar systems in which the environment follows the Ornstein--Uhlenbeck process in Eq.~(\ref{ou-langevin}). We set $m=0$ for both examples. Both systems describe production and removal dynamics of a single species. In the first example, production and removal rates are proportional to $\sigma$ when $\sigma>0$ and zero otherwise. In the second example the rates are each proportional to $|\sigma|$. These examples are not used in the main paper, we report them here for completeness, as they may prove useful for future applications of the $\tau$FE algorithm.

\subsection{Rates $R_{r,\sigma}(n) = \alpha_r  \sigma \Theta(\sigma)$ }

We look at the example $R_{r,\sigma}(n) = \alpha_r  \sigma \Theta(\sigma)$, where $\Theta(\sigma)$ is the Heaviside function, $\Theta(\sigma)=1$ for $\sigma>0$ and $\Theta(\sigma)=0$ otherwise. For $m=0$, we find
\begin{equation}
R_{r}^*(n) =\alpha_r  \dfrac{v}{\sqrt{\pi}},
\end{equation}
and
\begin{align*}
&\avg{\overline R_r(n) \overline R_s(n)}-R_{r}^*(n)R_{s}^*(n)=\\
&\quad \frac{\alpha_r \alpha_s v^2}{24 \pi  \Delta t^2} \Bigg\{\dfrac{1}{\lambda^2} \left[24 \pi  \left(\lambda \Delta t-1\right)-\pi ^2+ 12 \log ^2(2)\right]+ \\
&\quad \dfrac{4
   e^{-\lambda \Delta t}}{\lambda^2} \left[3 \left(6 \sqrt{e^{2 \lambda \Delta t}-1}+\pi \right)-4 e^{\lambda \Delta t} \log \left(\sqrt{e^{2 \lambda \Delta t}-1}+e^{\lambda \Delta t}\right)+6 \tan^{-1}\left(\frac{1}{\sqrt{e^{2 \lambda \Delta t}-1}}\right)\right] \\
   &\quad-\dfrac{32}{\lambda^2} \text{Re}\left(i \sin ^{-1}\left(e^{\lambda \Delta t}\right)\right)-6 \bigg[\dfrac{1}{\lambda^2}\log ^2\left(\sqrt{1-e^{-2 \lambda \Delta t}}+1\right)+\dfrac{4 \Delta t }{\lambda} \log \left(\sqrt{1-e^{-2 \lambda \Delta t}}+1\right) \\
& \quad-\dfrac{4 \Delta t \log (2)}{\lambda} - \dfrac{\log (4)}{\lambda^2}\log
   \left(\sqrt{1-e^{-2 \lambda \Delta t}}+1\right)-\dfrac{4 \Delta t}{\lambda}\tanh ^{-1}\left(e^{-\lambda \Delta t} \sqrt{e^{2 \lambda \Delta t}-1}\right) \\
   &\quad- \dfrac{2}{\lambda^2} \text{Li}_2\left(\frac{1}{2} \left(1-\sqrt{1-e^{-2 \lambda \Delta
   t}}\right)\right)+ \dfrac{\log ^2(2)}{\lambda^2} +2 \Delta t^2\bigg]-24\Bigg\}, \numberthis
\end{align*}
where Re($\cdot$) denotes the real part, and $\text{Li}_2(\cdot)$ is the polylogarithm of order 2.

\subsection{Rates $R_{r,\sigma}(n) = \alpha_r  |\sigma|$}

For this case (and setting again $m=0$), we find
\begin{equation}
R_{r}^*(n)= \alpha_r \dfrac{2 v}{\sqrt{\pi}},
\end{equation}
and
\begin{align*}
&\avg{\overline R_r(n) \overline R_s(n)}-R_{r}^*(\bn)R_{s}^*(n)=\\
& \quad \frac{\alpha_r \alpha_s v^2}{6 \pi  \Delta t^2} \Bigg\{\dfrac{12 \Delta t}{\lambda}\left[-2 \log \left(\sqrt{1-e^{-2 \lambda \Delta t}}+1\right)+2 \tanh
   ^{-1}\left(\sqrt{1-e^{-2 \lambda \Delta t}}\right)+\pi +\log (4)\right] \\
   & \quad +\dfrac{1}{\lambda^2}\Bigg[72 e^{- \lambda \Delta t } \sqrt{e^{2 \lambda \Delta t}-1}-6 \log ^2\left(\frac{1}{2} \left(\sqrt{1-e^{-2 \lambda \Delta t}}+1\right)\right)+24 e^{-\lambda \Delta t} \tan ^{-1}\left(\frac{1}{\sqrt{e^{2 \lambda \Delta t}-1}}\right) \\
   &\quad -48 \tanh ^{-1}\left(e^{-\lambda \Delta t} \sqrt{e^{2 \lambda \Delta t}-1}\right)+12
   \text{Li}_2\left(\frac{1}{2} \left(1-\sqrt{1-e^{-2 \lambda \Delta t}}\right)\right)-\pi ^2-12 \pi +12 \log ^2(2)\Bigg] \\
   &\quad -12 \left(\Delta t^2+2\right)\Bigg\}. \numberthis
\end{align*}

\end{appendix}

\end{document}